\documentclass[twocolumn, aps, prl, showpacs,preprintnumbers,bibnotes]{revtex4-1}
\pdfoutput=1
\usepackage[T1]{fontenc}

\usepackage{graphicx}
\usepackage{bm}
\usepackage{color}
\usepackage[T1]{fontenc}
\usepackage{amsmath}
\usepackage{upgreek}
\usepackage{nicefrac}
\usepackage{hyperref}
\usepackage{geometry}
\begin{document}

\title{Quantum state engineering of a Hubbard system with ultracold fermions}
\author{Christie S. Chiu}
\author{Geoffrey Ji}
\author{Anton Mazurenko}
\author{Daniel Greif}
\author{Markus Greiner}
\email{greiner@physics.harvard.edu}
\affiliation{
  Department of Physics, Harvard University,
  Cambridge, Massachusetts, 02138, USA}

\date{\today}

\begin{abstract}
Accessing new regimes in quantum simulation requires the development of new techniques for quantum state preparation.
We demonstrate the quantum state engineering of a strongly correlated many-body state of the two-component repulsive Fermi-Hubbard model on a square lattice.
Our scheme makes use of an ultralow entropy doublon band insulator created through entropy redistribution.
After isolating the band insulator, we change the underlying potential to expand it into a half-filled system.
The final many-body state realized shows strong antiferromagnetic correlations and a temperature below the exchange energy.
We observe an increase in entropy, which we find is likely caused by the many-body physics in the last step of the scheme.
This technique is promising for low-temperature studies of cold-atom-based lattice models.

\end{abstract}


\maketitle

Understanding and controlling complex many-body quantum physics is an important research frontier in quantum information, condensed matter physics, and quantum chemistry. 
Quantum simulation has emerged as a powerful tool for computing many-body quantum phases and dynamics, with the potential to exceed simulations on classical computers \cite{Feynman1982, Lloyd1996}.
By engineering highly coherent many-body systems, a wide variety of Hamiltonians can be studied \cite{Georgescu2014}. 
A unique platform for scalable quantum simulation is ultracold atoms, where the development of quantum gas microscopy has enabled control at the single atom level \cite{Bakr2010, Sherson2010}.
Quantum simulation extends to other promising platforms such as ion traps, superconducting circuits, solid state systems, Rydberg atoms, and photonic systems \cite{Bloch2012, Blatt2012, Aspuru-Guzik2012, Weimer2010, Devoret2013, Awschalom2013}. 

\begin{figure}[t!]
	\centering
	\includegraphics{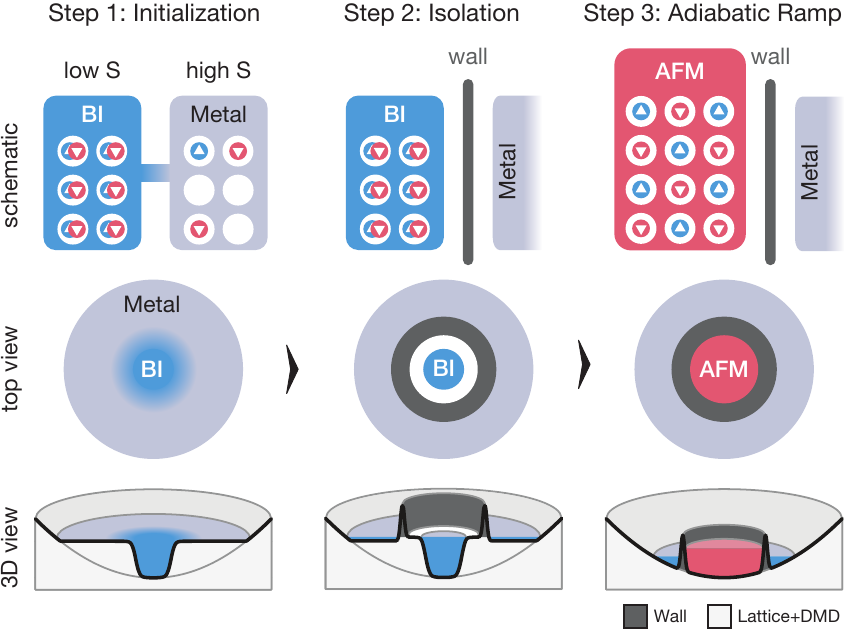}
	\caption{Illustration of the quantum state engineering scheme. (Top row) A low-density metallic state removes entropy from a band insulator (BI), after which the two states can be isolated thermally. The BI can then be ramped into an antiferromagnetic (AFM) state by increasing the number of available sites. (Middle row) Map of density inhomogeneity and states in our experimental setup. (Bottom row) We implement our scheme by engineering optical potential landscapes to change the Hamiltonian at each step (see main text and \cite{SI}).}
	\label{fig:overview}
\end{figure}

A major challenge of all these platforms is creating a coherent quantum many-body state, which is often the ground state.
Traditionally, cold atom experiments in optical lattices realize quantum states by loading an evaporatively cooled quantum gas into the lattice potential \cite{Bloch2008}.
This approach has been very successful \cite{Esslinger2010, Greif2013, Hart2015}, but the minimum achievable temperatures for fermionic systems are limited by reduced cooling efficiency at low temperatures.
An alternative approach is quantum state engineering. 
Generally, this method realizes an isolated pure quantum state by initializing one wavefunction under an initial Hamiltonian, then changing the Hamiltonian while preserving coherence during time evolution so that the accompanying wavefunction becomes the target state (see Fig.~\ref{fig:overview}).
Several platforms have used different versions of quantum state engineering to create desired quantum states \cite{Roos1999, Islam2011, Simon2011, Labuhn2014, Bernien2017}, and schemes have been proposed for ultracold fermionic atoms \cite{Bernier2009, Ho2009, Lubasch2011}. 
The site-resolved readout and control afforded by quantum gas microscope experiments \cite{Bakr2010, Sherson2010, Haller2015, Cheuk2016a, Mazurenko2017, Hilker2017, Edge2015, Miranda2015, Yamamoto2016, Brown2017} are perfect tools to implement quantum state engineering of many-body states of ultracold fermionic atoms in optical lattices.

Here we demonstrate quantum state engineering for a many-body state of fermionic atoms in the Hubbard model.
This model describes spin-1/2 fermions on a lattice with nearest-neighbor tunneling $t$ and repulsive on-site interaction $U$.
Under this model, a coexistence of phases can be realized through inhomogenous particle density in global thermal equilibrium \cite{Greif2016}.
A metal exists at low particle density, characterized by a large density of states and high entropy per particle.
At half-filling (one particle per site), an antiferromagnet emerges, where spins arrange in an alternating pattern.
This phase is gapped in the charge sector by $U$, but has nonzero density of states due to low-energy spin excitations.
The band insulator (BI) appears when the band is completely filled with two particles per site, and thus has a large energy gap equal to the bandgap, vanishing density of states, and vanishing entropy per particle.
Because of the differing density of states, under fixed global atom number and global entropy the density inhomogeneity can be engineered to produce low-entropy states.

If a BI and metal are in thermal contact, entropy flows from the BI into the metal, see Fig.~\ref{fig:overview}.
By using a fully gapped state, we optimize this entropy redistribution technique \cite{Mazurenko2017, Esslinger2010, Bernier2009, Ho2009}.
The result is an ultra-low entropy BI initial state with an entropy per particle as low as $0.016(3)\,k_\mathrm{B}$ in units of the Boltzmann constant, over an order of magnitude lower than the lowest value previously achieved with entropy redistribution \cite{Mazurenko2017}.
In the next step we thermally isolate the low-entropy region by suppressing particle transport between the BI and reservoir.
Finally, we convert the gapped BI into a strongly-correlated many-body state at half-filling.
This final state has a nearest-neighbor spin correlator of $C_1=-0.21(1)$ reflecting strong antiferromagnetic character and a temperature of $k_\mathrm{B}T/t=0.46(2)$.

Our experimental setup consists of a balanced spin mixture of the two lowest hyperfine states of fermionic $^6\mathrm{Li}$ in a combined square optical lattice and blue-detuned potential.
We set $t/h = 0.89(1)\,\mathrm{kHz}$ in units of the Planck constant and $U/t=7.7(3)$ or $U/t=5.9(2)$ \cite{SI}.
The quantum gas lies in the object plane of a quantum gas microscope \cite{SI}, allowing both atom imaging and potential control at the site-resolved level \cite{Parsons2015}.
Such precise control is achieved by placing two digital micromirror devices (DMD1 and DMD2) in the image plane and projecting their patterns with blue-detuned light \cite{SI}.
The DMD1 pattern is designed to engineer the coexistence of phases through changing the optical potential and therefore particle density across the sample, as in \cite{Mazurenko2017}.
DMD2 creates the isolating wall in the second step of our scheme.

\begin{figure}[t]
	\centering
	\includegraphics{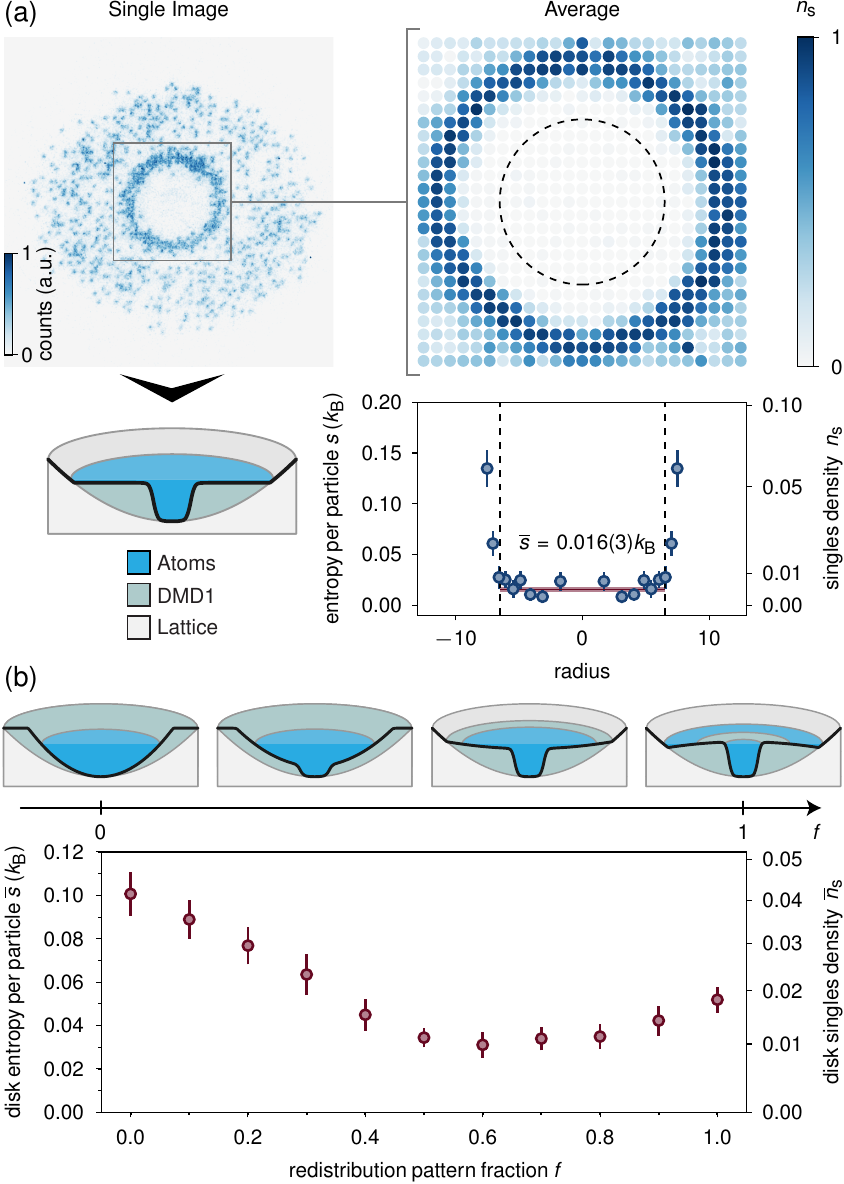}
	\caption{(color online). Ultra-low entropy BI at $U/t=7.7(3)$. (a) Raw fluorescence image of single BI,  optical potential schematic, and average density map of 50 BI realizations. Through entropy redistribution, we create a BI with $>130$ sites and average entropy per particle $0.016(3)\,k_{\mathrm{B}}$. Error bars denote standard error of 50 measurements and azimuthal averaging. (b) By continuously tuning the optical potential between a harmonic trap and our entropy redistribution pattern at constant BI size at $U/t=5.9(2)$, we see a decrease in BI entropy due to increased entropy redistribution efficiency. The final pattern yields a slightly higher entropy than the optimum, but is necessary for our scheme. Error bars denote standard error of >20 measurements each with 133 lattice sites.}
	\label{fig:step1}
\end{figure}

The success of quantum state engineering schemes fundamentally depends upon initial state preparation.
The initial density distribution consists of two regions of constant but different density: the doublon-filled center and the surrounding metallic reservoir (see Fig.~\ref{fig:step1}a), created with DMD1 by setting the potential offset between the two regions to $\Delta\approx2U$.
Following entropy redistribution we achieve an ultra-low entropy BI of more than $130$ sites.
Due to light-assisted collisions which occur during the imaging process, sites initially containing doublons appear as empty \cite{Parsons2015}.
We obtain the entropy per particle on a single site from the measured singles density $n_s$ \cite{SI}.
The average singles density $\overline{n}_s$ across the BI region is $0.4(1)\%$, corresponding to an upper bound for the average entropy per particle across the region of $0.016(3)\,k_\mathrm{B}$.
This signifies a 50-fold reduction in entropy compared to a homogenous system, showing that the technique is highly efficient \cite{SI}.
This entropy is significantly lower than that of the lowest-entropy two-component BIs realized in cold-atom systems thus far \cite{Greif2016, Cheuk2016a}.

Most cold atom experiments take place in a harmonic trap, where some entropy redistribution is already present because of inhomogenous particle density. 
We compare entropy redistribution efficiencies between a harmonic pattern and the employed pattern for quantum state engineering by interpolating linearly between these two profiles, parameterized by the fraction $f$.
Atom number, total entropy, and BI size are kept constant.
As shown in Fig.~\ref{fig:step1}b, entropy redistribution reduces the BI entropy per particle by more than a factor of 3 compared to the harmonic trap, even for a pattern which has not been optimized for redistribution efficiency as in Fig.~\ref{fig:step1}a.
We find a slight increase in entropy moving to the final pattern, which may result from a denser reservoir or from a loss of thermal contact between the system and reservoir.
Indeed, the $f=1$ pattern exhibits a ring of zero density between these two regions.
This ring is necessary in the next step of the quantum state engineering scheme.

\begin{figure}[b]
	\centering
	\includegraphics{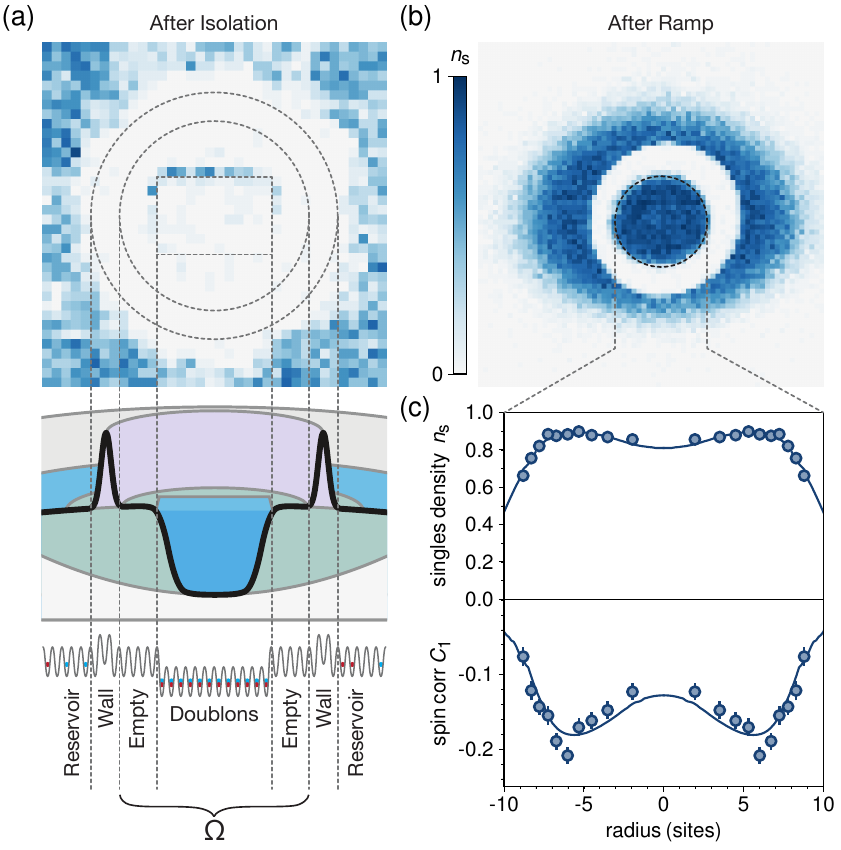}\\
	\caption{(color online). (a) Average density map (40 realizations) and configuration of entropy redistribution at $U/t=5.9(2)$ after isolation. Imperfections in the optical potential manifest as singly-occupied sites, as seen at the upper edge of the box. (b) Average density map of 41 images after ramp highlighting how the insulating wall separates the inner and outer regions, with initialization via disk pattern. (c) Corresponding density and nearest-neighbor correlator profiles vs. radius after ramp. The nearest-neighbor correlations are antiferromagnetic with a strength of up to $C_1=-0.21(1)$. A simultaneous fit to both profiles (solid line) gives a temperature of $k_\mathrm{B}T/t=0.46(2)$. The fit is limited to radius $9$, to avoid effects from the insulating wall. Error bars denote standard error of >40 sets of correlation maps and azimuthal averaging. For (b)+(c), $U/t = 7.7(3)$.}
	\label{fig:step2}
\end{figure}

After initializing the low-entropy BI, the next step is to isolate it from the remaining atoms. 
We adjust the entropy redistribution pattern such that the BI is surrounded by holes, see Fig.~\ref{fig:step2}a. 
To ensure full isolation of the BI, we subsequently raise a circular wall with a thickness of about $3$ sites using DMD2 \cite{SI}. 
We set the wall diameter to a value larger than the BI size. 
The region $\Omega$ inside the wall therefore contains both the BI and empty sites. 
For the shape of the BI we choose either a circular $12$-site diameter disk (similar to the $f=1$ configuration) or a rectangular 8-site by 12-site box.
Both regions in $\Omega$ are approximately homogenous in density with energy offset $\Delta\approx2U$.
To ensure the two regions have the desired densities, we set the global chemical potential inside $\Omega$ to a value below $\Delta$ by adjusting the total atom number \cite{SI}. 

For the box-shaped configuration (see Fig.~\ref{fig:step2}a), the entropy per particle within $\Omega$ is $0.25(1)\,k_\mathrm{B}$.
This entropy is greater than that of the pure BI because it includes both doublon and hole regions; indeed, the pure BI entropy per particle away from the box edge is only $0.08(1)\,k_\mathrm{B}$, so the greatest entropy contribution to $\Omega$ is from the boundary between the regions.
More specifically, if the box potential is not perfectly aligned with the lattice sites, the potential offset on sites close to the edge can be modified.
Even if the box is aligned, the microscope point spread function smooths the potential across one or two lattice sites.
These effects lead to density defects on both sides of the box edge that are visible as singly-occupied sites and increased entropy, see for example the upper box edge in Fig.~\ref{fig:step2}a.
The ring-shaped wall has a negligible effect on initial entropy, confirmed through comparing the entropy with and without the wall.

The final step in our quantum state engineering scheme is to convert the initial state into the target many-body state.
For this measurement we use the disk pattern for the BI to reduce alignment sensitivity.
To ensure half-filling in the final state, the wall diameter is set such that the number of holes and doublons within $\Omega$ is approximately equal. 
After initialization and isolation we slowly remove the potential offset between holes and doublons by reducing the DMD1 laser power \cite{SI}. 
In Fig.~\ref{fig:step2}b we show the measured singles density $n_s$ after a $40\,$ms linear ramp of the potential offset.
The atomic density extends over the entire region $\Omega$ and sharply decreases at the inner edge of the wall, indicating particle transport has occurred from the doublon core to the surrounding empty sites.
The inner and outer regions are separated by the insulating wall, marked by a ring of empty sites.
In the final state, atoms in $\Omega$ are expected to show antiferromagnetic correlations, whose strength reflect the adiabaticity of the ramp.
The nearest-neighbor spin correlations, measured with a technique established in previous work \cite{Parsons2016}, are strongly antiferromagnetic with values up to $C_1=4\langle\hat{S}^z_i\hat{S}^z_{i+1}\rangle=-0.21(1)$. Here $\hat{S}^z_i$ denotes the standard spin-$1/2$ operator along the $z$-direction on site $i$.
These observations demonstrate a successful implementation of quantum state engineering, where a strongly-correlated many-body state is created from an initially uncorrelated BI of doublons. 

Locally changing density and spin correlations within $\Omega$ originate from the underlying harmonic confinement created by the lattice lasers.
The maximum in the singles density radial profile indicates the density is above half-filling in the center and continuously decreases for larger radii, see Fig.~\ref{fig:step2}c.
We intentionally keep this confinement to study whether the system is in thermal equilibrium.
When applying a simultaneous fit of exact theoretical predictions to both measured radial profiles with shared fit parameters, we find reasonable agreement \cite{Khatami2011, Parsons2016}.
This shows that the system within $\Omega$ is consistent with a thermal equilibrium state.
In order to determine whether deviations from thermal equilibrium or finite-size effects are present, more detailed knowledge of corrections to the exact confinement potential is required. 

From the fit we obtain a temperature in $\Omega$ of $k_\mathrm{B}T/t=0.46(2)$, which is comparable to the temperatures achieved so far in harmonic traps \cite{Brown2017}, but still higher than the lowest value of $k_\mathrm{B}T/t=0.25(2)$ achieved with entropy redistribution \cite{Mazurenko2017}.
Although this temperature is surprisingly low given the simple ramp scheme used here, the system is still far from the ground state.
Besides the nonzero entropy of the initial BI, this nonzero temperature may result from non-adiabaticity of the ramp or residual heating.
We now explore both possibilities.

\begin{figure}[b]
	\centering
	\includegraphics{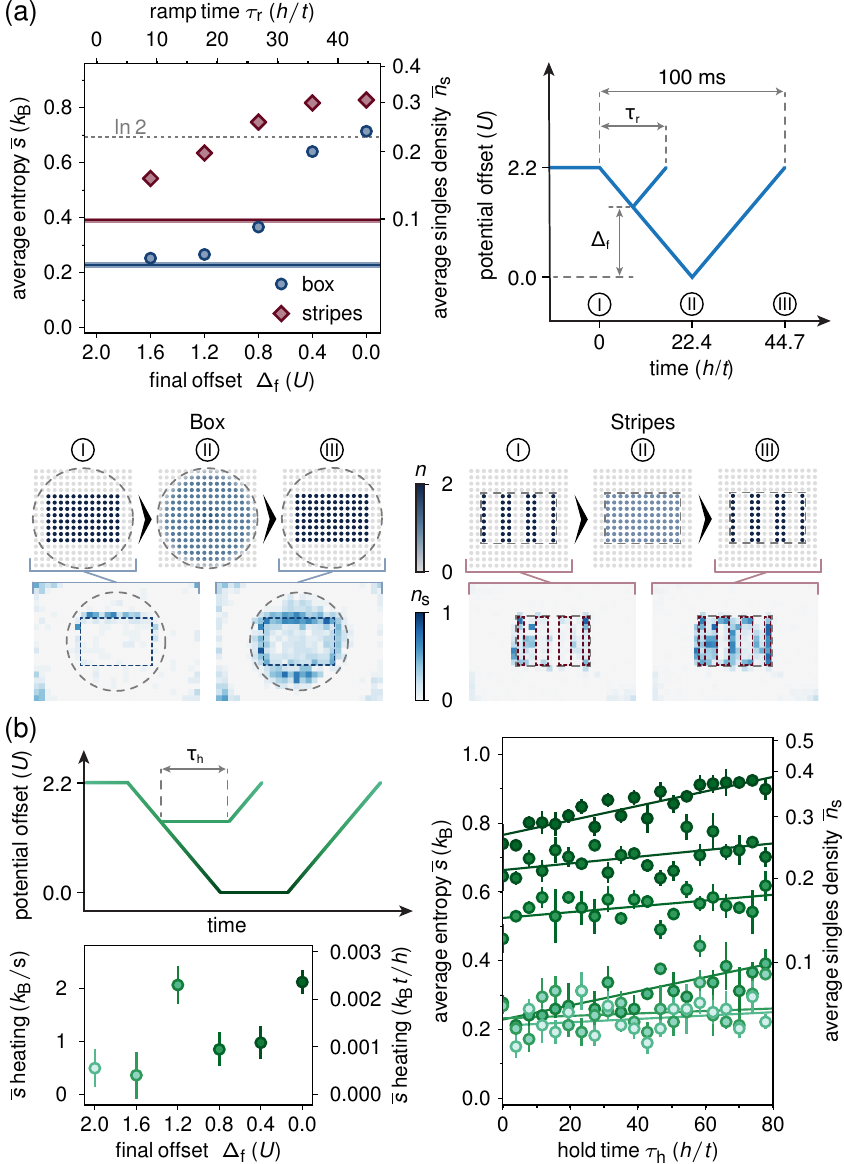}\\
	\caption{(color online). Examining final ramp adiabaticity at $U/t=5.9(3)$. (a) A round trip measurement beginning with an isolated box of doublons surrounded by holes demonstrates non-adiabaticity predominantly in the second half of the offset ramp-off (circles). Adiabaticity is not significantly improved by initializing the holes and doublons in stripes (diamonds). Horizontal lines with shading indicate reference measurements and uncertainty, taken with no ramp. Lower panels show schematic images for the particle density $n$ and measured average singles density maps for the box (left) and stripe (right) configurations at different times throughout the round-trip ramp. Dashed lines indicate the wall inner edge, while dotted lines enclose BI regions. Error bars are smaller than the markers, and denote standard error of 40 (187) measurements for the box (stripe) pattern. (b) We quantify heating rates at various points throughout the ramp (upper left), which enable us to approximate the contribution of entropy increase due to heating. Error bars for entropy measurements (right) denote standard error of 5 measurements; error bars for heating rates (lower left) are from the fits.}
	\label{fig:step3}
\end{figure}

We first study non-adiabaticity by examining the entropy increase after completing and reversing the offset ramp \cite{SI}.
For this measurement we use the box pattern for the lowest initial entropy in $\Omega$.
As heating effects are negligible in the initial state, a perfectly adiabatic process implies measuring the same entropy as this initial state.
When varying the endpoint of the ramp $\Delta_f$, we find that the entropy per particle increases steadily as the ramp endpoint decreases, see Fig.~\ref{fig:step3}a.
The qualitative shape of the curve suggests a lack of adiabaticity largely throughout the second half of the ramp.
In this regime, $\Delta \approx U$ and particles can freely tunnel out of the doublon core.
For the full two-way ramp, we find an entropy increase of $0.46(2)\,k_\mathrm{B}$.
Although this increase strongly indicates a non-adiabatic ramp, it may actually be caused by greater heating rates during the ramp, for example due to changes in the many-body energy spectrum.

To distinguish heating during the ramp from non-adiabaticity, we measure the heating rate for each ramp endpoint by holding for a variable time $\tau_h$ before reversing the ramp and measuring the resulting entropy \cite{SI}.
Heating rates are generally greater than the initial heating rate, with values up to $2.1(2)\,k_\mathrm{B}/\mathrm{s}$, see Fig.~\ref{fig:step3}b.
The observed increase in heating rate at $\Delta \approx U$ indicates a drastic change in the many-body energy spectrum, as already suggested by the non-adiabaticity measurement of Fig.~\ref{fig:step3}a.
From the measured rates, we estimate an entropy increase from heating of $0.06\,k_\mathrm{B}$ for the full ramp.
This indicates that the majority of entropy increase does not originate from heating, but rather from non-adiabaticity.
When decreasing the ramp rate for the full two-way ramp, the final entropy increases, indicating that any improvement in adiabaticity is insufficient to overcome heating during the additional ramp time.

Despite the non-adiabaticity, achievement of low temperatures with such a simple ramp scheme is encouraging.
A possible improvement is to reduce the amount of required particle transport, which may reduce the non-adiabaticity.
We repeat the adiabaticity measurement for an initial system consisting of alternating stripes of holes and doublons surrounded by a box-shaped wall, see bottom right panel of Fig.~\ref{fig:step3}a.
While the initial entropy is worse than that of the box due to the more complex potential landscape, crucially this configuration yields no significant improvement in entropy increase.
This suggests that the dominant reason for non-adiabaticity lies within the many-body physics occurring during the ramp, which strongly depends on how the ramp is implemented and what intermediate phases are crossed \cite{Fubini2007, Cincio2009, Hoang2016}.
An improvement could be to avoid a closing charge gap in the many-body spectrum during the ramp, possibly by using a double-well superlattice.
Such a configuration has been predicted to be very efficient in numerical simulations \cite{Lubasch2011, Kantian2016}. 

In conclusion, we have implemented a quantum state engineering scheme to create a fermionic many-body state.
Through adjusting the initial balance of doubly-occupied and unoccupied sites, this technique offers the flexibility to vary the doping of the sample on the single-atom level.
Furthermore, the remarkably low initial entropies afforded by entropy redistribution may enable even lower temperatures at arbitrary doping to search for signatures of a $d$-wave superfluid state \cite{Lee2006}.
However, additional studies must be conducted to determine the optimum path in parameter space which minimizes the entropy.
Analogous quantum state engineering schemes can be designed for studies of stripe phases with strongly magnetic atoms, massively entangled spin states, and adiabatic quantum computation \cite{Mazloom2017, Zhang2013, Farhi2001}.

\begin{acknowledgments}
	We would like to thank Ehud Altman, Lawrence Cheuk, Andrew Daley, Frederik G\"org, Adam Kaufman, Maddy Kaufman, Julian Leonard, Ahmed Omran, Philipp Preiss, Achim Rosch, Eric Tai, and Hao Zhang for insightful discussions.
	We acknowledge support from AFOSR (MURI), ARO (MURI, NDSEG), the Gordon and Betty Moore foundation EPiQS initiative, NSF (CUA, GRFP), and SNSF.
\end{acknowledgments}

\clearpage

\newgeometry{left=0.7in,right=0.7in,top=0.7in,bottom=0.7in}
	
\onecolumngrid

\begin{center} 
	\begin{Large} \textbf{Supplemental material} \end{Large}
\end{center}
\medskip
\medskip

\setcounter{section}{0}
\setcounter{subsection}{0}
\setcounter{figure}{0}
\setcounter{equation}{0}
\setcounter{NAT@ctr}{0}

\makeatletter
\renewcommand{\bibnumfmt}[1]{[S#1]}
\renewcommand{\citenumfont}[1]{S#1}
\renewcommand{\thefigure}{S\@arabic\c@figure} 
\makeatother
\renewcommand{\theequation}{S\arabic{equation}} 

\twocolumngrid

\section{Experimental Sequence}
\begin{figure}[b]
	\centering
	\includegraphics[width=0.3\textwidth]{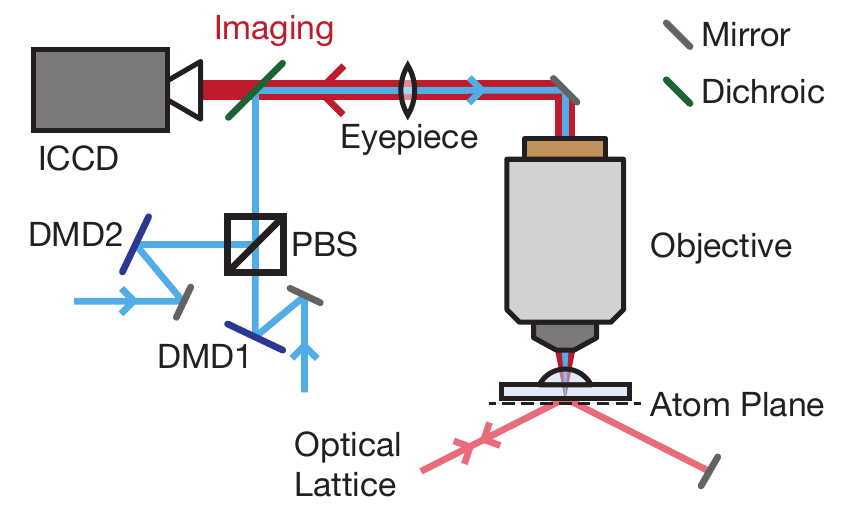}\\
	\caption{Schematic view of the optical setup. We implement our scheme with a quantum gas microscope and two digital micromirror devices DMD1 and DMD2, which enable control of the optical potential at the site-resolved level.}
	\label{sifig:setup}
\end{figure}

\begin{figure}[t]
	\centering
	\includegraphics[width=0.49\textwidth]{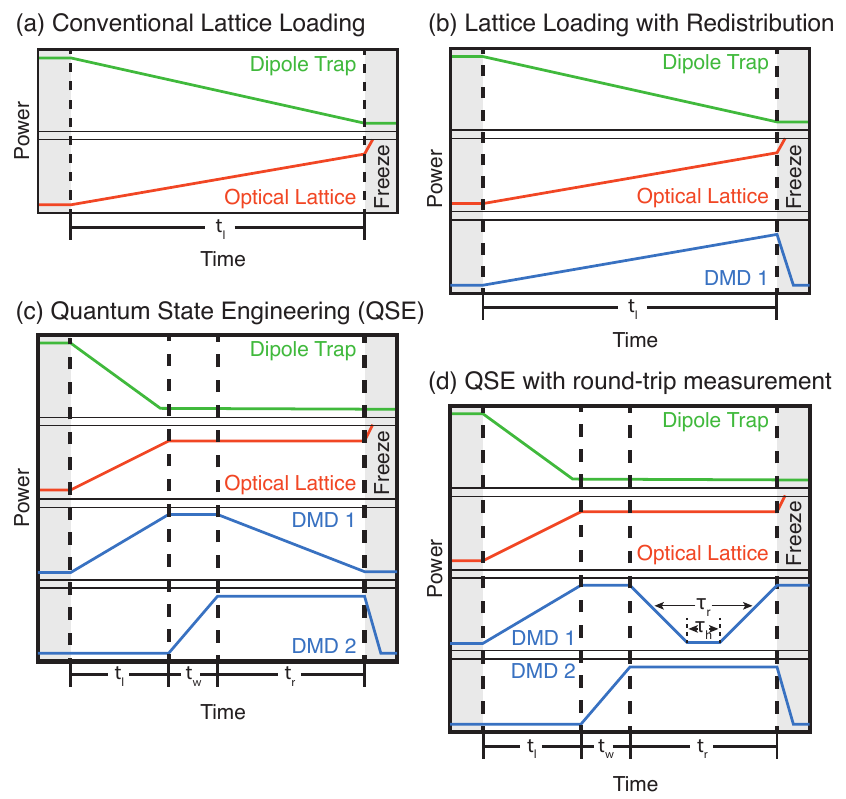}
	\caption{
		Schematic views of various lattice loading sequences, where $t_l$ is the time for lattice loading, $t_w$ is the time for ramping on the isolation wall, and $t_r$ is the offset ramp time. (a) Lattice loading sequence as used in \cite{Parsons2016a}. (b) Sequence as used in \cite{Mazurenko2017a} and data in Figure 2 of the main text. (c) Sequence for the quantum state engineering protocol, as used in Figure 3. Note the addition of a second independent DMD, DMD2, required to implement the scheme. (d) Sequence to measure the round-trip entropy increase and heating rates during the final ramp of the quantum state engineering protocol, as used in Figure 4 of the main text. The times $\tau_r$ and $\tau_h$ are as defined in the main text.
	}
	\label{sifig:ramps}
\end{figure}

All experiments start with a sample of evaporatively-cooled Lithium-6 in a balanced spin mix of the energetically lowest two hyperfine levels. The sample is confined in the axial direction into a single plane of a red-detuned optical lattice and in the radial direction by an optical dipole trap, as in previous works. The lattice and both DMDs are then turned on in various sequences while the initial confinement is shut off, depending on the experimental goal. A schematic view of the experimental setup is presented in Figure \ref{sifig:setup}, and a schematic view of the various sequences is presented in Figure \ref{sifig:ramps}. In the data presented in Figure 2 of the main text, the lattice loading time $t_l$ is $100\,\mathrm{ms}$. In the data presented in Figures 3 and 4 of the main text, $t_l$ is $200\,\mathrm{ms}$, the isolation wall ramp time $t_w$ is $20\,\mathrm{ms}$, and the offset ramp time $t_r$ is dependent on the specific experiment.

\section{Hubbard parameters}

We set the depth of the optical lattice to $7.5(1)\,E_{\mathrm{R}}$, where the recoil energy $E_{\mathrm{R}}/h = 25.6\,\mathrm{kHz}$ and $h$ is the Planck constant, such that the bare tunneling between neighboring sites in the square lattice is $t/h=0.89(1)\,\mathrm{kHz}$.
We use the broad Feshbach resonance at $832\,\mathrm{G}$ to set the scattering length to either $a=129\,a_0$ or $a=210\,a_0$, in units of the Bohr radius \cite{Zurn2013}.
The interaction parameter is then $U/t=7.7(3)$ or $U/t=5.9(2)$.

\subsection{Calibration of the Hubbard parameters}

We calculate the Hubbard $t$ parameter by measuring the excitation energy from the ground band to the first excited band via lattice modulation, and fitting to a band structure calculated from the Mathieu equation. The Hubbard $U$ parameters used in the main text are determined by preparing Hubbard systems at two different temperatures and performing a simultaneous fit of the temperature, chemical potential, interaction energy $U$, and overall harmonic confinement frequency to the singles density and nearest-neighbor spin correlator using results from a numerical linked cluster expansion (NLCE) algorithm \cite{Khatami2011a}.

\section{Entropy calculation}

In the limit of a perfect BI, the measured on-site singles density $n_s$ is small. We assume there are no holes, and we have a spin balance. Then on any given site we have a spin up (down) particle with probability $n_s/2$ ($n_s/2$), and a doublon with probability $1-n_s$. The entropy per particle on a given site $s$ is then given in units of the Boltzmann constant $k_\mathrm{B}$ by dividing the total entropy $S$ by the total atom number $N$, of one site across many realizations, all based on the probability $p_i$ of having some outcome $i$ and number of particles $n_i$ of that outcome:
\begin{align}
s &\equiv S / N = \frac{-k_\mathrm{B} \sum_i p_i \ln{p_i}}{\sum_i p_i n_i}\nonumber \\
&= \frac1{2}\left[- n_s \ln{\left(n_s / 2\right)} - (1-n_s) \ln{\left(1-n_s\right)} \right] k_\mathrm{B}
\end{align}
This entropy is an upper bound because correlations between different sites lower the average entropy, but are not included in this estimate. We have assumed the number of particles per site is exactly 2, which introduces an error of (maximally) $0.002\,k_\mathrm{B}$. This method of calculating the entropy upper bound is used for all entropy values calculated from the data shown in Figure 2 of the main text. Since the regions that are averaged over in these data are homogenous and close to the BI phase, we can calculate $\overline{s}$, the average entropy per particle across a portion of the system, by using Equation S1 but replacing $n_s$ with the average singles density across the portion $\overline{n}_s$. 

On the other hand, for the entropy values calculated from data in Figures 3 and 4 of the main text, the regions that are averaged over include sites that are expected to be empty. These sites have the same relationship between $n_s$ and $S$ as sites that are almost full, but the number of particles is now close to 0 rather than 2. The regions therefore have on average 1 particle per site since the enclosed regions have the same number of doublon sites as hole sites. We assume an average of exactly 1 particle per site, which from an inequality in number of doublon sites versus hole sites of up to 5 sites results in a maximal relative error of $2.5\%$. We calculate the average entropy by measuring $\overline{n}_s$ across the entire region, which results in an upper bound for the entropy since entropy is a concave function with respect to singles density. We perform the region averaging to decrease the effect of the bias of the entropy estimator, which decreases as the number of samples increases. For $n\approx100$ as in the presented data, this bias is maximally $0.006\,k_\mathrm{B}$. The entropy per particle is therefore calculated as:
\begin{align}
\overline{s} &\equiv \overline{S} / \overline{N}\nonumber \\
&= \left[- \overline{n}_s \ln{\left(\overline{n}_s / 2\right)} - (1-\overline{n}_s) \ln{\left(1-\overline{n}_s\right)} \right] k_\mathrm{B}
\end{align}
Note that this result differs from equation S1 by a factor of two due to the difference in particle number.

Uncertainties on entropy per particle values quoted in the main text are derived from standard error propagation of the standard deviation of the measured singles density.

\section{Entropy redistribution efficiency}

For fixed global entropy $\overline{S}$ and global particle number $\overline{N}$, we can approximate the entropy redistribution efficiency based on the difference in entropy per particle of two regions by some factor $\alpha$:

\begin{equation}
\left( \frac{\overline{S}}{\overline{N}} \right) _\mathrm{res} \equiv \alpha \left( \frac{\overline{S}}{\overline{N}} \right) _\mathrm{center} 
\end{equation}

Then the total entropy per particle is given by the total entropy of the system, divided by the total particle number:

\begin{align}
\left( \frac{\overline{S}}{\overline{N}} \right) _\mathrm{total} &= \frac{\overline{S}_\mathrm{res} + \overline{S}_\mathrm{center}}{\overline{N}_\mathrm{res} + \overline{N}_\mathrm{center}}\\
&= \frac{\left( \frac{\overline{S}}{\overline{N}} \right) _\mathrm{res} \overline{N}_\mathrm{res}+\left( \frac{\overline{S}}{\overline{N}} \right) _\mathrm{center} \overline{N}_\mathrm{center}}{\overline{N}_\mathrm{res} + \overline{N}_\mathrm{center}}\\
&= \frac{\alpha \left( \frac{\overline{S}}{\overline{N}} \right) _\mathrm{center} \overline{N}_\mathrm{res} + \left( \frac{\overline{S}}{\overline{N}} \right) _\mathrm{center} \overline{N}_\mathrm{center}}{\overline{N}_\mathrm{res} + \overline{N}_\mathrm{center}}\\
&= \left( \frac{\overline{S}}{\overline{N}} \right) _\mathrm{center} \frac{\alpha + \overline{N}_\mathrm{center}/\overline{N}_\mathrm{res}}{1+  \overline{N}_\mathrm{center}/\overline{N}_\mathrm{res}}
\end{align}

Notice now that in the limit where the number of atoms in the reservoir is much larger than the number of atoms in the center system, the center entropy is suppressed from the total entropy by a factor of $\alpha$.

Given a global entropy per particle in the experiment of approximately $1.0\,k_{\mathrm{B}}$, we find a relative entropy reduction of $\alpha\approx50$ for this entropy redistribution scheme. 

\begin{figure}[tb]
	\centering
	\includegraphics{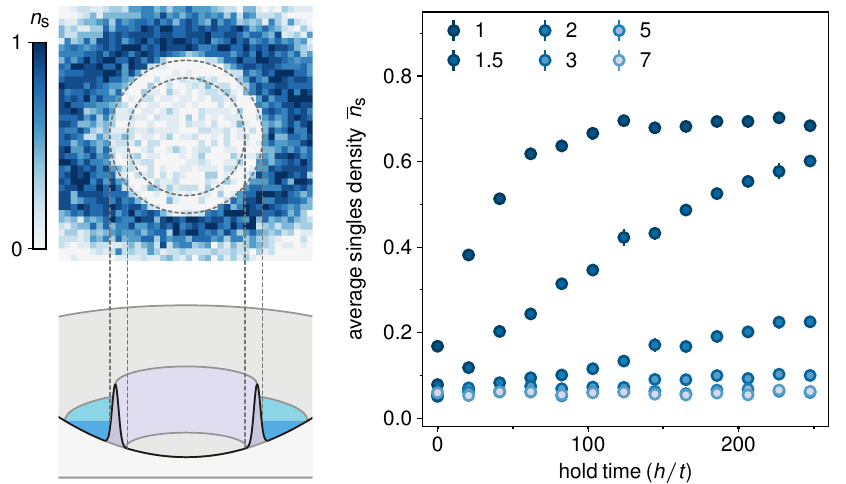}\\
	\caption{Verifying wall insulation. After preparing a state containing fewer than 3 atoms, we raise up a circular wall as in the quantum state engineering protocol. The potential used to prepare the initial state is ramped off, leaving an atom distribution (upper left) and attractive harmonic well potential (lower left). The rate of particle transport into the central region is measured, for varying wall thicknesses in lattice sites (right). For experiments discussed in the main text, we use a wall thickness of 3 lattice sites.}
	\label{sifig:verif}
\end{figure}

\begin{figure*}[t]
	\centering
	
	\begin{minipage}[c]{0.65\textwidth}
		\includegraphics[width=\textwidth]{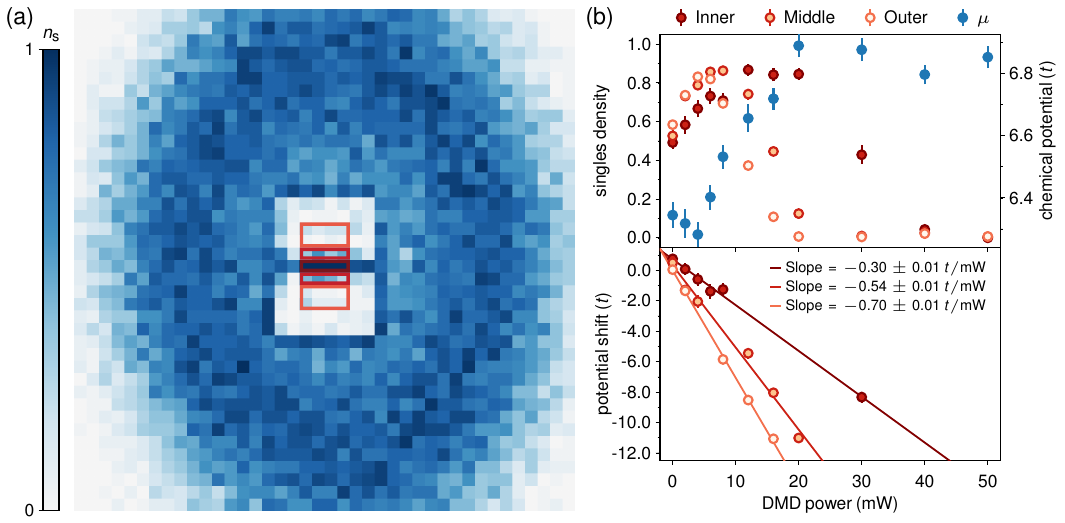}
	\end{minipage}\hfill
	\begin{minipage}[c]{0.34\textwidth}
		\caption{
			Calibrating DMD potential. (a) Singles density at a DMD power of $16\,\mathrm{mW}$. The structure in the center is a result of the DMD pattern, and the overall shape of the atom distribution is from the underlying harmonic confinement. As the power is increased, the density in the boxed regions changes. (b) Upper: Plot of the density in the boxed regions versus DMD power. The fitted global chemical potential, $\mu$, is plotted along the right axis, and increases greatest when the filling decreases in the outer boxed region. Lower: Fitted potential shift, determined from the singles density, the equation of state, and the fitted global chemical potential.
		}
		\label{sifig:dmdcalib}
	\end{minipage}
\end{figure*}

\section{Parameter verification for quantum state engineering protocol}

The quantum state engineering scheme described in the main text requires several well-tuned parameters which are not discussed in the main text. In this section we briefly show that the chosen parameters are optimized or otherwise sufficient for implementation of our scheme.

First, we ensure that the wall isolates the central system from its reservoir. To do this, we study particle flow into a system designed to contain only empty sites. The entropy redistribution pattern is modified to be completely deconfining in the center, such that the wall encloses a system with fewer than 3 particles. The outer radius of the wall is kept constant at 12 sites. Upon removal of the deconfining potential, the center becomes an attractive harmonic well, see left panel of Fig. \ref{sifig:verif}. If the wall is not insulating, particles can tunnel into the system. We study the rate at which particles enter, for various thicknesses of the confining wall, as seen in the right panel of Fig. \ref{sifig:verif}.

As expected, the rate of inward particle flow decreases with wall thickness. As several virtual tunneling processes to high energy states are required for a particle to tunnel across the wall, we expect the effective wall tunneling time to grow exponentially with the wall thickness. Experimentally, however, the wall cannot be made arbitrarily wide, because it must occupy a region containing no particles. Increasing the size of this region decreases the size of the usable region $\Omega$. We note instead that the wall need only be sufficiently insulating on the timescale of the adiabatic ramp into the final many-body state. Therefore, we find the minimum wall thickness for which on average the entropy increases by less than about $0.05\,k_\mathrm{B}$, given a conservative ramp time of $40\,\mathrm{ms}$. This corresponds to a wall 3 sites wide.

Second, because imaging does not distinguish doublons from holes, we explicitly check that doublons are loaded into the desired regions of our initial configuration. We determine the minimum required potential offset by increasing the offset from zero and observing the atom density of the box increase past half-filling and into the BI regime. The potential offset between the sites with holes and sites is doublons is then set to a value greater than this minimum value.

\subsection{Digital micromirror device potential calibration}

To calibrate the potential created by the DMD, we apply a pattern consisting of two boxes separated by a narrow strip of the same width used in the single-site wide regions in Fig.~4a of the main text. The resulting density distribution for an intermediate DMD power is shown in Fig.~\ref{sifig:dmdcalib}b, with the areas used for calibration indicated by the boxed regions. By changing the optical power incident on the DMD and determining the average density distribution, we determine the effect of the DMD on the site occupation, see Fig.~\ref{sifig:dmdcalib}b, upper panel. We also fit the density distributions of the atoms outside of the area affected by the DMD to determine the global chemical potential and temperature.

We then use theoretical calculations using a NLCE algorithm to invert the site-dependent singles occupations and extract local chemical potentials \cite{Khatami2011a}. Values above the maximum theoretical singles occupation and within two standard errors of zero density are discarded. The global chemical potentials are subtracted from these values to determine the potential difference caused by the DMD. The offsets are fit to a line whose slope is the potential offset per unit DMD power, see Fig.~\ref{sifig:dmdcalib}b, lower panel. All data presented in the main text were taken using the maximum DMD power of $50\,\mathrm{mW}$, where the offset between adjacent sites within and outside the potential (i.e. the inner and middle region) is $-2.0(2)\,\mathrm{U}$, and the maximum offset is $-6.0(2)\,\mathrm{U}$, where $U$ is the smaller interaction energy used in the main text.


\begin{thebibliography}{44}%
	\makeatletter
	\providecommand \@ifxundefined [1]{%
		\@ifx{#1\undefined}
	}%
	\providecommand \@ifnum [1]{%
		\ifnum #1\expandafter \@firstoftwo
		\else \expandafter \@secondoftwo
		\fi
	}%
	\providecommand \@ifx [1]{%
		\ifx #1\expandafter \@firstoftwo
		\else \expandafter \@secondoftwo
		\fi
	}%
	\providecommand \natexlab [1]{#1}%
	\providecommand \enquote  [1]{``#1''}%
	\providecommand \bibnamefont  [1]{#1}%
	\providecommand \bibfnamefont [1]{#1}%
	\providecommand \citenamefont [1]{#1}%
	\providecommand \href@noop [0]{\@secondoftwo}%
	\providecommand \href [0]{\begingroup \@sanitize@url \@href}%
	\providecommand \@href[1]{\@@startlink{#1}\@@href}%
	\providecommand \@@href[1]{\endgroup#1\@@endlink}%
	\providecommand \@sanitize@url [0]{\catcode `\\12\catcode `\$12\catcode
		`\&12\catcode `\#12\catcode `\^12\catcode `\_12\catcode `\%12\relax}%
	\providecommand \@@startlink[1]{}%
	\providecommand \@@endlink[0]{}%
	\providecommand \url  [0]{\begingroup\@sanitize@url \@url }%
	\providecommand \@url [1]{\endgroup\@href {#1}{\urlprefix }}%
	\providecommand \urlprefix  [0]{URL }%
	\providecommand \Eprint [0]{\href }%
	\providecommand \doibase [0]{http://dx.doi.org/}%
	\providecommand \selectlanguage [0]{\@gobble}%
	\providecommand \bibinfo  [0]{\@secondoftwo}%
	\providecommand \bibfield  [0]{\@secondoftwo}%
	\providecommand \translation [1]{[#1]}%
	\providecommand \BibitemOpen [0]{}%
	\providecommand \bibitemStop [0]{}%
	\providecommand \bibitemNoStop [0]{.\EOS\space}%
	\providecommand \EOS [0]{\spacefactor3000\relax}%
	\providecommand \BibitemShut  [1]{\csname bibitem#1\endcsname}%
	\let\auto@bib@innerbib\@empty
	\bibitem [{\citenamefont {Feynman}(1982)}]{Feynman1982}%
	\BibitemOpen
	\bibfield  {author} {\bibinfo {author} {\bibfnamefont {R.~P.}\ \bibnamefont
			{Feynman}},\ }\href {\doibase 10.1007/BF02650179} {\bibfield  {journal}
		{\bibinfo  {journal} {International Journal of Theoretical Physics}\ }\textbf
		{\bibinfo {volume} {21}},\ \bibinfo {pages} {467} (\bibinfo {year}
		{1982})}\BibitemShut {NoStop}%
	\bibitem [{\citenamefont {Lloyd}(1996)}]{Lloyd1996}%
	\BibitemOpen
	\bibfield  {author} {\bibinfo {author} {\bibfnamefont {S.}~\bibnamefont
			{Lloyd}},\ }\href {\doibase 10.1126/science.273.5278.1073} {\bibfield
		{journal} {\bibinfo  {journal} {Science}\ }\textbf {\bibinfo {volume}
			{273}},\ \bibinfo {pages} {1073} (\bibinfo {year} {1996})}\BibitemShut
	{NoStop}%
	\bibitem [{\citenamefont {Georgescu}\ \emph {et~al.}(2014)\citenamefont
		{Georgescu}, \citenamefont {Ashhab},\ and\ \citenamefont
		{Nori}}]{Georgescu2014}%
	\BibitemOpen
	\bibfield  {author} {\bibinfo {author} {\bibfnamefont {I.~M.}\ \bibnamefont
			{Georgescu}}, \bibinfo {author} {\bibfnamefont {S.}~\bibnamefont {Ashhab}}, \
		and\ \bibinfo {author} {\bibfnamefont {F.}~\bibnamefont {Nori}},\ }\href
	{\doibase 10.1103/RevModPhys.86.153} {\bibfield  {journal} {\bibinfo
			{journal} {Reviews of Modern Physics}\ }\textbf {\bibinfo {volume} {86}},\
		\bibinfo {pages} {153} (\bibinfo {year} {2014})}\BibitemShut {NoStop}%
	\bibitem [{\citenamefont {Bakr}\ \emph {et~al.}(2010)\citenamefont {Bakr},
		\citenamefont {Peng}, \citenamefont {Tai}, \citenamefont {Ma}, \citenamefont
		{Simon}, \citenamefont {Gillen}, \citenamefont {Foelling}, \citenamefont
		{Pollet},\ and\ \citenamefont {Greiner}}]{Bakr2010}%
	\BibitemOpen
	\bibfield  {author} {\bibinfo {author} {\bibfnamefont {W.~S.}\ \bibnamefont
			{Bakr}}, \bibinfo {author} {\bibfnamefont {A.}~\bibnamefont {Peng}}, \bibinfo
		{author} {\bibfnamefont {M.~E.}\ \bibnamefont {Tai}}, \bibinfo {author}
		{\bibfnamefont {R.}~\bibnamefont {Ma}}, \bibinfo {author} {\bibfnamefont
			{J.}~\bibnamefont {Simon}}, \bibinfo {author} {\bibfnamefont {J.~I.}\
			\bibnamefont {Gillen}}, \bibinfo {author} {\bibfnamefont {S.}~\bibnamefont
			{Foelling}}, \bibinfo {author} {\bibfnamefont {L.}~\bibnamefont {Pollet}}, \
		and\ \bibinfo {author} {\bibfnamefont {M.}~\bibnamefont {Greiner}},\ }\href
	{\doibase 10.1126/science.1192368} {\bibfield  {journal} {\bibinfo  {journal}
			{Science}\ }\textbf {\bibinfo {volume} {329}},\ \bibinfo {pages} {547}
		(\bibinfo {year} {2010})}\BibitemShut {NoStop}%
	\bibitem [{\citenamefont {Sherson}\ \emph {et~al.}(2010)\citenamefont
		{Sherson}, \citenamefont {Weitenberg}, \citenamefont {Endres}, \citenamefont
		{Cheneau}, \citenamefont {Bloch},\ and\ \citenamefont {Kuhr}}]{Sherson2010}%
	\BibitemOpen
	\bibfield  {author} {\bibinfo {author} {\bibfnamefont {J.~F.}\ \bibnamefont
			{Sherson}}, \bibinfo {author} {\bibfnamefont {C.}~\bibnamefont {Weitenberg}},
		\bibinfo {author} {\bibfnamefont {M.}~\bibnamefont {Endres}}, \bibinfo
		{author} {\bibfnamefont {M.}~\bibnamefont {Cheneau}}, \bibinfo {author}
		{\bibfnamefont {I.}~\bibnamefont {Bloch}}, \ and\ \bibinfo {author}
		{\bibfnamefont {S.}~\bibnamefont {Kuhr}},\ }\href {\doibase
		10.1038/nature09378} {\bibfield  {journal} {\bibinfo  {journal} {Nature}\
		}\textbf {\bibinfo {volume} {467}},\ \bibinfo {pages} {68} (\bibinfo {year}
		{2010})}\BibitemShut {NoStop}%
	\bibitem [{\citenamefont {Bloch}\ \emph {et~al.}(2012)\citenamefont {Bloch},
		\citenamefont {Dalibard},\ and\ \citenamefont
		{Nascimb{\`{e}}ne}}]{Bloch2012}%
	\BibitemOpen
	\bibfield  {author} {\bibinfo {author} {\bibfnamefont {I.}~\bibnamefont
			{Bloch}}, \bibinfo {author} {\bibfnamefont {J.}~\bibnamefont {Dalibard}}, \
		and\ \bibinfo {author} {\bibfnamefont {S.}~\bibnamefont {Nascimb{\`{e}}ne}},\
	}\href {\doibase 10.1038/NPHYS2259} {\bibfield  {journal} {\bibinfo
			{journal} {Nature Physics}\ }\textbf {\bibinfo {volume} {8}},\ \bibinfo
		{pages} {267} (\bibinfo {year} {2012})}\BibitemShut {NoStop}%
	\bibitem [{\citenamefont {Blatt}\ and\ \citenamefont {Roos}(2012)}]{Blatt2012}%
	\BibitemOpen
	\bibfield  {author} {\bibinfo {author} {\bibfnamefont {R.}~\bibnamefont
			{Blatt}}\ and\ \bibinfo {author} {\bibfnamefont {C.~F.}\ \bibnamefont
			{Roos}},\ }\href {\doibase 10.1038/NPHYS2252} {\bibfield  {journal} {\bibinfo
			{journal} {Nature Physics}\ }\textbf {\bibinfo {volume} {8}},\ \bibinfo
		{pages} {277} (\bibinfo {year} {2012})}\BibitemShut {NoStop}%
	\bibitem [{\citenamefont {Aspuru-Guzik}\ and\ \citenamefont
		{Walther}(2012)}]{Aspuru-Guzik2012}%
	\BibitemOpen
	\bibfield  {author} {\bibinfo {author} {\bibfnamefont {A.}~\bibnamefont
			{Aspuru-Guzik}}\ and\ \bibinfo {author} {\bibfnamefont {P.}~\bibnamefont
			{Walther}},\ }\href {\doibase 10.1038/NPHYS2253} {\bibfield  {journal}
		{\bibinfo  {journal} {Nature Physics}\ }\textbf {\bibinfo {volume} {8}},\
		\bibinfo {pages} {285} (\bibinfo {year} {2012})}\BibitemShut {NoStop}%
	\bibitem [{\citenamefont {Weimer}\ \emph {et~al.}(2010)\citenamefont {Weimer},
		\citenamefont {M{\"{u}}ller}, \citenamefont {Lesanovsky}, \citenamefont
		{Zoller},\ and\ \citenamefont {B{\"{u}}chler}}]{Weimer2010}%
	\BibitemOpen
	\bibfield  {author} {\bibinfo {author} {\bibfnamefont {H.}~\bibnamefont
			{Weimer}}, \bibinfo {author} {\bibfnamefont {M.}~\bibnamefont
			{M{\"{u}}ller}}, \bibinfo {author} {\bibfnamefont {I.}~\bibnamefont
			{Lesanovsky}}, \bibinfo {author} {\bibfnamefont {P.}~\bibnamefont {Zoller}},
		\ and\ \bibinfo {author} {\bibfnamefont {H.~P.}\ \bibnamefont
			{B{\"{u}}chler}},\ }\href {\doibase 10.1038/NPHYS1614} {\bibfield  {journal}
		{\bibinfo  {journal} {Nature Physics}\ }\textbf {\bibinfo {volume} {6}},\
		\bibinfo {pages} {382} (\bibinfo {year} {2010})}\BibitemShut {NoStop}%
	\bibitem [{\citenamefont {Devoret}\ and\ \citenamefont
		{Schoelkopf}(2013)}]{Devoret2013}%
	\BibitemOpen
	\bibfield  {author} {\bibinfo {author} {\bibfnamefont {M.~H.}\ \bibnamefont
			{Devoret}}\ and\ \bibinfo {author} {\bibfnamefont {R.~J.}\ \bibnamefont
			{Schoelkopf}},\ }\href {\doibase 10.1126/science.1231930} {\bibfield
		{journal} {\bibinfo  {journal} {Science}\ }\textbf {\bibinfo {volume}
			{339}},\ \bibinfo {pages} {1169} (\bibinfo {year} {2013})}\BibitemShut
	{NoStop}%
	\bibitem [{\citenamefont {Awschalom}\ \emph {et~al.}(2013)\citenamefont
		{Awschalom}, \citenamefont {Bassett}, \citenamefont {Dzurak}, \citenamefont
		{Hu},\ and\ \citenamefont {Petta}}]{Awschalom2013}%
	\BibitemOpen
	\bibfield  {author} {\bibinfo {author} {\bibfnamefont {D.~D.}\ \bibnamefont
			{Awschalom}}, \bibinfo {author} {\bibfnamefont {L.~C.}\ \bibnamefont
			{Bassett}}, \bibinfo {author} {\bibfnamefont {A.~S.}\ \bibnamefont {Dzurak}},
		\bibinfo {author} {\bibfnamefont {E.~L.}\ \bibnamefont {Hu}}, \ and\ \bibinfo
		{author} {\bibfnamefont {J.~R.}\ \bibnamefont {Petta}},\ }\href {\doibase
		10.1126/science.1231364} {\bibfield  {journal} {\bibinfo  {journal}
			{Science}\ }\textbf {\bibinfo {volume} {339}},\ \bibinfo {pages} {1174}
		(\bibinfo {year} {2013})}\BibitemShut {NoStop}%
	\bibitem [{\citenamefont {Bloch}\ \emph {et~al.}(2008)\citenamefont {Bloch},
		\citenamefont {Dalibard},\ and\ \citenamefont {Zwerger}}]{Bloch2008}%
	\BibitemOpen
	\bibfield  {author} {\bibinfo {author} {\bibfnamefont {I.}~\bibnamefont
			{Bloch}}, \bibinfo {author} {\bibfnamefont {J.}~\bibnamefont {Dalibard}}, \
		and\ \bibinfo {author} {\bibfnamefont {W.}~\bibnamefont {Zwerger}},\ }\href
	{\doibase 10.1103/RevModPhys.80.885} {\bibfield  {journal} {\bibinfo
			{journal} {Reviews of Modern Physics}\ }\textbf {\bibinfo {volume} {80}},\
		\bibinfo {pages} {885} (\bibinfo {year} {2008})}\BibitemShut {NoStop}%
	\bibitem [{\citenamefont {Esslinger}(2010)}]{Esslinger2010}%
	\BibitemOpen
	\bibfield  {author} {\bibinfo {author} {\bibfnamefont {T.}~\bibnamefont
			{Esslinger}},\ }\href {\doibase 10.1146/annurev-conmatphys-070909-104059}
	{\bibfield  {journal} {\bibinfo  {journal} {Annual Reviews of Condensed
				Matter Physics}\ }\textbf {\bibinfo {volume} {1}},\ \bibinfo {pages} {129}
		(\bibinfo {year} {2010})}\BibitemShut {NoStop}%
	\bibitem [{\citenamefont {Greif}\ \emph {et~al.}(2013)\citenamefont {Greif},
		\citenamefont {Uehlinger}, \citenamefont {Jotzu}, \citenamefont {Tarruell},\
		and\ \citenamefont {Esslinger}}]{Greif2013}%
	\BibitemOpen
	\bibfield  {author} {\bibinfo {author} {\bibfnamefont {D.}~\bibnamefont
			{Greif}}, \bibinfo {author} {\bibfnamefont {T.}~\bibnamefont {Uehlinger}},
		\bibinfo {author} {\bibfnamefont {G.}~\bibnamefont {Jotzu}}, \bibinfo
		{author} {\bibfnamefont {L.}~\bibnamefont {Tarruell}}, \ and\ \bibinfo
		{author} {\bibfnamefont {T.}~\bibnamefont {Esslinger}},\ }\href {\doibase
		10.1126/science.1236362} {\bibfield  {journal} {\bibinfo  {journal}
			{Science}\ }\textbf {\bibinfo {volume} {340}},\ \bibinfo {pages} {1307}
		(\bibinfo {year} {2013})}\BibitemShut {NoStop}%
	\bibitem [{\citenamefont {Hart}\ \emph {et~al.}(2015)\citenamefont {Hart},
		\citenamefont {Duarte}, \citenamefont {Yang}, \citenamefont {Liu},
		\citenamefont {Paiva}, \citenamefont {Khatami}, \citenamefont {Scalettar},
		\citenamefont {Trivedi}, \citenamefont {Huse},\ and\ \citenamefont
		{Hulet}}]{Hart2015}%
	\BibitemOpen
	\bibfield  {author} {\bibinfo {author} {\bibfnamefont {R.~A.}\ \bibnamefont
			{Hart}}, \bibinfo {author} {\bibfnamefont {P.~M.}\ \bibnamefont {Duarte}},
		\bibinfo {author} {\bibfnamefont {T.-L.}\ \bibnamefont {Yang}}, \bibinfo
		{author} {\bibfnamefont {X.}~\bibnamefont {Liu}}, \bibinfo {author}
		{\bibfnamefont {T.}~\bibnamefont {Paiva}}, \bibinfo {author} {\bibfnamefont
			{E.}~\bibnamefont {Khatami}}, \bibinfo {author} {\bibfnamefont {R.~T.}\
			\bibnamefont {Scalettar}}, \bibinfo {author} {\bibfnamefont {N.}~\bibnamefont
			{Trivedi}}, \bibinfo {author} {\bibfnamefont {D.~A.}\ \bibnamefont {Huse}}, \
		and\ \bibinfo {author} {\bibfnamefont {R.~G.}\ \bibnamefont {Hulet}},\ }\href
	{\doibase 10.1038/nature14223} {\bibfield  {journal} {\bibinfo  {journal}
			{Nature}\ }\textbf {\bibinfo {volume} {519}},\ \bibinfo {pages} {211}
		(\bibinfo {year} {2015})}\BibitemShut {NoStop}%
	\bibitem [{\citenamefont {Roos}\ \emph {et~al.}(1999)\citenamefont {Roos},
		\citenamefont {Zeiger}, \citenamefont {Rohde}, \citenamefont {N{\"{a}}gerl},
		\citenamefont {Eschner}, \citenamefont {Leibfried}, \citenamefont
		{Schmidt-Kaler},\ and\ \citenamefont {Blatt}}]{Roos1999}%
	\BibitemOpen
	\bibfield  {author} {\bibinfo {author} {\bibfnamefont {C.}~\bibnamefont
			{Roos}}, \bibinfo {author} {\bibfnamefont {T.}~\bibnamefont {Zeiger}},
		\bibinfo {author} {\bibfnamefont {H.}~\bibnamefont {Rohde}}, \bibinfo
		{author} {\bibfnamefont {H.~C.}\ \bibnamefont {N{\"{a}}gerl}}, \bibinfo
		{author} {\bibfnamefont {J.}~\bibnamefont {Eschner}}, \bibinfo {author}
		{\bibfnamefont {D.}~\bibnamefont {Leibfried}}, \bibinfo {author}
		{\bibfnamefont {F.}~\bibnamefont {Schmidt-Kaler}}, \ and\ \bibinfo {author}
		{\bibfnamefont {R.}~\bibnamefont {Blatt}},\ }\href {\doibase
		10.1103/PhysRevLett.83.4713} {\bibfield  {journal} {\bibinfo  {journal}
			{Physical Review Letters}\ }\textbf {\bibinfo {volume} {83}},\ \bibinfo
		{pages} {4713} (\bibinfo {year} {1999})}\BibitemShut {NoStop}%
	\bibitem [{\citenamefont {Islam}\ \emph {et~al.}(2011)\citenamefont {Islam},
		\citenamefont {Edwards}, \citenamefont {Kim}, \citenamefont {Korenblit},
		\citenamefont {Noh}, \citenamefont {Carmichael}, \citenamefont {Lin},
		\citenamefont {Duan}, \citenamefont {{Joseph Wang}}, \citenamefont
		{Freericks},\ and\ \citenamefont {Monroe}}]{Islam2011}%
	\BibitemOpen
	\bibfield  {author} {\bibinfo {author} {\bibfnamefont {R.}~\bibnamefont
			{Islam}}, \bibinfo {author} {\bibfnamefont {E.}~\bibnamefont {Edwards}},
		\bibinfo {author} {\bibfnamefont {K.}~\bibnamefont {Kim}}, \bibinfo {author}
		{\bibfnamefont {S.}~\bibnamefont {Korenblit}}, \bibinfo {author}
		{\bibfnamefont {C.}~\bibnamefont {Noh}}, \bibinfo {author} {\bibfnamefont
			{H.}~\bibnamefont {Carmichael}}, \bibinfo {author} {\bibfnamefont {G.-D.}\
			\bibnamefont {Lin}}, \bibinfo {author} {\bibfnamefont {L.-M.}\ \bibnamefont
			{Duan}}, \bibinfo {author} {\bibfnamefont {C.-C.}\ \bibnamefont {{Joseph
					Wang}}}, \bibinfo {author} {\bibfnamefont {J.}~\bibnamefont {Freericks}}, \
		and\ \bibinfo {author} {\bibfnamefont {C.}~\bibnamefont {Monroe}},\ }\href
	{\doibase 10.1038/ncomms1374} {\bibfield  {journal} {\bibinfo  {journal}
			{Nature Communications}\ }\textbf {\bibinfo {volume} {2}},\ \bibinfo {pages}
		{377} (\bibinfo {year} {2011})}\BibitemShut {NoStop}%
	\bibitem [{\citenamefont {Simon}\ \emph {et~al.}(2011)\citenamefont {Simon},
		\citenamefont {Bakr}, \citenamefont {Ma}, \citenamefont {Tai}, \citenamefont
		{Preiss},\ and\ \citenamefont {Greiner}}]{Simon2011}%
	\BibitemOpen
	\bibfield  {author} {\bibinfo {author} {\bibfnamefont {J.}~\bibnamefont
			{Simon}}, \bibinfo {author} {\bibfnamefont {W.~S.}\ \bibnamefont {Bakr}},
		\bibinfo {author} {\bibfnamefont {R.}~\bibnamefont {Ma}}, \bibinfo {author}
		{\bibfnamefont {M.~E.}\ \bibnamefont {Tai}}, \bibinfo {author} {\bibfnamefont
			{P.~M.}\ \bibnamefont {Preiss}}, \ and\ \bibinfo {author} {\bibfnamefont
			{M.}~\bibnamefont {Greiner}},\ }\href {\doibase 10.1038/nature09994}
	{\bibfield  {journal} {\bibinfo  {journal} {Nature}\ }\textbf {\bibinfo
			{volume} {472}},\ \bibinfo {pages} {307} (\bibinfo {year}
		{2011})}\BibitemShut {NoStop}%
	\bibitem [{\citenamefont {Labuhn}\ \emph {et~al.}(2014)\citenamefont {Labuhn},
		\citenamefont {Ravets}, \citenamefont {Barredo}, \citenamefont
		{B{\'{e}}guin}, \citenamefont {Nogrette}, \citenamefont {Lahaye},\ and\
		\citenamefont {Browaeys}}]{Labuhn2014}%
	\BibitemOpen
	\bibfield  {author} {\bibinfo {author} {\bibfnamefont {H.}~\bibnamefont
			{Labuhn}}, \bibinfo {author} {\bibfnamefont {S.}~\bibnamefont {Ravets}},
		\bibinfo {author} {\bibfnamefont {D.}~\bibnamefont {Barredo}}, \bibinfo
		{author} {\bibfnamefont {L.}~\bibnamefont {B{\'{e}}guin}}, \bibinfo {author}
		{\bibfnamefont {F.}~\bibnamefont {Nogrette}}, \bibinfo {author}
		{\bibfnamefont {T.}~\bibnamefont {Lahaye}}, \ and\ \bibinfo {author}
		{\bibfnamefont {A.}~\bibnamefont {Browaeys}},\ }\href {\doibase
		10.1103/PhysRevA.90.023415} {\bibfield  {journal} {\bibinfo  {journal}
			{Physical Review A}\ }\textbf {\bibinfo {volume} {90}},\ \bibinfo {pages}
		{023415} (\bibinfo {year} {2014})}\BibitemShut {NoStop}%
	\bibitem [{\citenamefont {Bernien}\ \emph {et~al.}(2017)\citenamefont
		{Bernien}, \citenamefont {Schwartz}, \citenamefont {Keesling}, \citenamefont
		{Levine}, \citenamefont {Omran}, \citenamefont {Pichler}, \citenamefont
		{Choi}, \citenamefont {Zibrov}, \citenamefont {Endres}, \citenamefont
		{Greiner}, \citenamefont {Vuleti{\'{c}}},\ and\ \citenamefont
		{Lukin}}]{Bernien2017}%
	\BibitemOpen
	\bibfield  {author} {\bibinfo {author} {\bibfnamefont {H.}~\bibnamefont
			{Bernien}}, \bibinfo {author} {\bibfnamefont {S.}~\bibnamefont {Schwartz}},
		\bibinfo {author} {\bibfnamefont {A.}~\bibnamefont {Keesling}}, \bibinfo
		{author} {\bibfnamefont {H.}~\bibnamefont {Levine}}, \bibinfo {author}
		{\bibfnamefont {A.}~\bibnamefont {Omran}}, \bibinfo {author} {\bibfnamefont
			{H.}~\bibnamefont {Pichler}}, \bibinfo {author} {\bibfnamefont
			{S.}~\bibnamefont {Choi}}, \bibinfo {author} {\bibfnamefont {A.~S.}\
			\bibnamefont {Zibrov}}, \bibinfo {author} {\bibfnamefont {M.}~\bibnamefont
			{Endres}}, \bibinfo {author} {\bibfnamefont {M.}~\bibnamefont {Greiner}},
		\bibinfo {author} {\bibfnamefont {V.}~\bibnamefont {Vuleti{\'{c}}}}, \ and\
		\bibinfo {author} {\bibfnamefont {M.~D.}\ \bibnamefont {Lukin}},\ }\href
	{\doibase 10.1038/nature24622} {\bibfield  {journal} {\bibinfo  {journal}
			{Nature}\ }\textbf {\bibinfo {volume} {551}},\ \bibinfo {pages} {579}
		(\bibinfo {year} {2017})}\BibitemShut {NoStop}%
	\bibitem [{\citenamefont {Bernier}\ \emph {et~al.}(2009)\citenamefont
		{Bernier}, \citenamefont {Kollath}, \citenamefont {Georges}, \citenamefont
		{{De Leo}}, \citenamefont {Gerbier}, \citenamefont {Salomon},\ and\
		\citenamefont {K{\"{o}}hl}}]{Bernier2009}%
	\BibitemOpen
	\bibfield  {author} {\bibinfo {author} {\bibfnamefont {J.-S.}\ \bibnamefont
			{Bernier}}, \bibinfo {author} {\bibfnamefont {C.}~\bibnamefont {Kollath}},
		\bibinfo {author} {\bibfnamefont {A.}~\bibnamefont {Georges}}, \bibinfo
		{author} {\bibfnamefont {L.}~\bibnamefont {{De Leo}}}, \bibinfo {author}
		{\bibfnamefont {F.}~\bibnamefont {Gerbier}}, \bibinfo {author} {\bibfnamefont
			{C.}~\bibnamefont {Salomon}}, \ and\ \bibinfo {author} {\bibfnamefont
			{M.}~\bibnamefont {K{\"{o}}hl}},\ }\href {\doibase
		10.1103/PhysRevA.79.061601} {\bibfield  {journal} {\bibinfo  {journal}
			{Physical Review A}\ }\textbf {\bibinfo {volume} {79}},\ \bibinfo {pages}
		{061601(R)} (\bibinfo {year} {2009})}\BibitemShut {NoStop}%
	\bibitem [{\citenamefont {Ho}\ and\ \citenamefont {Zhou}()}]{Ho2009}%
	\BibitemOpen
	\bibfield  {author} {\bibinfo {author} {\bibfnamefont {T.-L.}\ \bibnamefont
			{Ho}}\ and\ \bibinfo {author} {\bibfnamefont {Q.}~\bibnamefont {Zhou}},\
	}\href@noop {} {\ }\Eprint {http://arxiv.org/abs/0911.5506v1}
	{arXiv:0911.5506v1} \BibitemShut {NoStop}%
	\bibitem [{\citenamefont {Lubasch}\ \emph {et~al.}(2011)\citenamefont
		{Lubasch}, \citenamefont {Murg}, \citenamefont {Schneider}, \citenamefont
		{Cirac},\ and\ \citenamefont {Ba{\~{n}}uls}}]{Lubasch2011}%
	\BibitemOpen
	\bibfield  {author} {\bibinfo {author} {\bibfnamefont {M.}~\bibnamefont
			{Lubasch}}, \bibinfo {author} {\bibfnamefont {V.}~\bibnamefont {Murg}},
		\bibinfo {author} {\bibfnamefont {U.}~\bibnamefont {Schneider}}, \bibinfo
		{author} {\bibfnamefont {J.~I.}\ \bibnamefont {Cirac}}, \ and\ \bibinfo
		{author} {\bibfnamefont {M.-C.~C.}\ \bibnamefont {Ba{\~{n}}uls}},\ }\href
	{\doibase 10.1103/PhysRevLett.107.165301} {\bibfield  {journal} {\bibinfo
			{journal} {Physical Review Letters}\ }\textbf {\bibinfo {volume} {107}},\
		\bibinfo {pages} {165301} (\bibinfo {year} {2011})}\BibitemShut {NoStop}%
	\bibitem [{\citenamefont {Haller}\ \emph {et~al.}(2015)\citenamefont {Haller},
		\citenamefont {Hudson}, \citenamefont {Kelly}, \citenamefont {Cotta},
		\citenamefont {Peaudecerf}, \citenamefont {Bruce},\ and\ \citenamefont
		{Kuhr}}]{Haller2015}%
	\BibitemOpen
	\bibfield  {author} {\bibinfo {author} {\bibfnamefont {E.}~\bibnamefont
			{Haller}}, \bibinfo {author} {\bibfnamefont {J.}~\bibnamefont {Hudson}},
		\bibinfo {author} {\bibfnamefont {A.}~\bibnamefont {Kelly}}, \bibinfo
		{author} {\bibfnamefont {D.~A.}\ \bibnamefont {Cotta}}, \bibinfo {author}
		{\bibfnamefont {B.}~\bibnamefont {Peaudecerf}}, \bibinfo {author}
		{\bibfnamefont {G.~D.}\ \bibnamefont {Bruce}}, \ and\ \bibinfo {author}
		{\bibfnamefont {S.}~\bibnamefont {Kuhr}},\ }\href {\doibase
		10.1038/NPHYS3403} {\bibfield  {journal} {\bibinfo  {journal} {Nature
				Physics}\ }\textbf {\bibinfo {volume} {11}},\ \bibinfo {pages} {738}
		(\bibinfo {year} {2015})}\BibitemShut {NoStop}%
	\bibitem [{\citenamefont {Cheuk}\ \emph {et~al.}(2016)\citenamefont {Cheuk},
		\citenamefont {Nichols}, \citenamefont {Lawrence}, \citenamefont {Okan},
		\citenamefont {Zhang}, \citenamefont {Khatami}, \citenamefont {Trivedi},
		\citenamefont {Paiva}, \citenamefont {Rigol},\ and\ \citenamefont
		{Zwierlein}}]{Cheuk2016a}%
	\BibitemOpen
	\bibfield  {author} {\bibinfo {author} {\bibfnamefont {L.~W.}\ \bibnamefont
			{Cheuk}}, \bibinfo {author} {\bibfnamefont {M.~A.}\ \bibnamefont {Nichols}},
		\bibinfo {author} {\bibfnamefont {K.~R.}\ \bibnamefont {Lawrence}}, \bibinfo
		{author} {\bibfnamefont {M.}~\bibnamefont {Okan}}, \bibinfo {author}
		{\bibfnamefont {H.}~\bibnamefont {Zhang}}, \bibinfo {author} {\bibfnamefont
			{E.}~\bibnamefont {Khatami}}, \bibinfo {author} {\bibfnamefont
			{N.}~\bibnamefont {Trivedi}}, \bibinfo {author} {\bibfnamefont
			{T.}~\bibnamefont {Paiva}}, \bibinfo {author} {\bibfnamefont
			{M.}~\bibnamefont {Rigol}}, \ and\ \bibinfo {author} {\bibfnamefont {M.~W.}\
			\bibnamefont {Zwierlein}},\ }\href {\doibase 10.1126/science.aag3349}
	{\bibfield  {journal} {\bibinfo  {journal} {Science}\ }\textbf {\bibinfo
			{volume} {353}},\ \bibinfo {pages} {1260} (\bibinfo {year}
		{2016})}\BibitemShut {NoStop}%
	\bibitem [{\citenamefont {Mazurenko}\ \emph {et~al.}(2017)\citenamefont
		{Mazurenko}, \citenamefont {Chiu}, \citenamefont {Ji}, \citenamefont
		{Parsons}, \citenamefont {Kan{\'{a}}sz-Nagy}, \citenamefont {Schmidt},
		\citenamefont {Grusdt}, \citenamefont {Demler}, \citenamefont {Greif},\ and\
		\citenamefont {Greiner}}]{Mazurenko2017}%
	\BibitemOpen
	\bibfield  {author} {\bibinfo {author} {\bibfnamefont {A.}~\bibnamefont
			{Mazurenko}}, \bibinfo {author} {\bibfnamefont {C.~S.}\ \bibnamefont {Chiu}},
		\bibinfo {author} {\bibfnamefont {G.}~\bibnamefont {Ji}}, \bibinfo {author}
		{\bibfnamefont {M.~F.}\ \bibnamefont {Parsons}}, \bibinfo {author}
		{\bibfnamefont {M.}~\bibnamefont {Kan{\'{a}}sz-Nagy}}, \bibinfo {author}
		{\bibfnamefont {R.}~\bibnamefont {Schmidt}}, \bibinfo {author} {\bibfnamefont
			{F.}~\bibnamefont {Grusdt}}, \bibinfo {author} {\bibfnamefont
			{E.}~\bibnamefont {Demler}}, \bibinfo {author} {\bibfnamefont
			{D.}~\bibnamefont {Greif}}, \ and\ \bibinfo {author} {\bibfnamefont
			{M.}~\bibnamefont {Greiner}},\ }\href {\doibase 10.1038/nature22362}
	{\bibfield  {journal} {\bibinfo  {journal} {Nature}\ }\textbf {\bibinfo
			{volume} {545}},\ \bibinfo {pages} {462} (\bibinfo {year}
		{2017})}\BibitemShut {NoStop}%
	\bibitem [{\citenamefont {Hilker}\ \emph {et~al.}(2017)\citenamefont {Hilker},
		\citenamefont {Salomon}, \citenamefont {Grusdt}, \citenamefont {Omran},
		\citenamefont {Boll}, \citenamefont {Demler}, \citenamefont {Bloch},\ and\
		\citenamefont {Gross}}]{Hilker2017}%
	\BibitemOpen
	\bibfield  {author} {\bibinfo {author} {\bibfnamefont {T.~A.}\ \bibnamefont
			{Hilker}}, \bibinfo {author} {\bibfnamefont {G.}~\bibnamefont {Salomon}},
		\bibinfo {author} {\bibfnamefont {F.}~\bibnamefont {Grusdt}}, \bibinfo
		{author} {\bibfnamefont {A.}~\bibnamefont {Omran}}, \bibinfo {author}
		{\bibfnamefont {M.}~\bibnamefont {Boll}}, \bibinfo {author} {\bibfnamefont
			{E.}~\bibnamefont {Demler}}, \bibinfo {author} {\bibfnamefont
			{I.}~\bibnamefont {Bloch}}, \ and\ \bibinfo {author} {\bibfnamefont
			{C.}~\bibnamefont {Gross}},\ }\href {\doibase 10.1126/science.aam8990}
	{\bibfield  {journal} {\bibinfo  {journal} {Science}\ }\textbf {\bibinfo
			{volume} {357}},\ \bibinfo {pages} {484} (\bibinfo {year}
		{2017})}\BibitemShut {NoStop}%
	\bibitem [{\citenamefont {Edge}\ \emph {et~al.}(2015)\citenamefont {Edge},
		\citenamefont {Anderson}, \citenamefont {Jervis}, \citenamefont {McKay},
		\citenamefont {Day}, \citenamefont {Trotzky},\ and\ \citenamefont
		{Thywissen}}]{Edge2015}%
	\BibitemOpen
	\bibfield  {author} {\bibinfo {author} {\bibfnamefont {G.~J.~A.}\
			\bibnamefont {Edge}}, \bibinfo {author} {\bibfnamefont {R.}~\bibnamefont
			{Anderson}}, \bibinfo {author} {\bibfnamefont {D.}~\bibnamefont {Jervis}},
		\bibinfo {author} {\bibfnamefont {D.~C.}\ \bibnamefont {McKay}}, \bibinfo
		{author} {\bibfnamefont {R.}~\bibnamefont {Day}}, \bibinfo {author}
		{\bibfnamefont {S.}~\bibnamefont {Trotzky}}, \ and\ \bibinfo {author}
		{\bibfnamefont {J.~H.}\ \bibnamefont {Thywissen}},\ }\href {\doibase
		10.1103/PhysRevA.92.063406} {\bibfield  {journal} {\bibinfo  {journal}
			{Physical Review A}\ }\textbf {\bibinfo {volume} {92}},\ \bibinfo {pages}
		{063406} (\bibinfo {year} {2015})}\BibitemShut {NoStop}%
	\bibitem [{\citenamefont {Miranda}\ \emph {et~al.}(2015)\citenamefont
		{Miranda}, \citenamefont {Inoue}, \citenamefont {Okuyama}, \citenamefont
		{Nakamoto},\ and\ \citenamefont {Kozuma}}]{Miranda2015}%
	\BibitemOpen
	\bibfield  {author} {\bibinfo {author} {\bibfnamefont {M.}~\bibnamefont
			{Miranda}}, \bibinfo {author} {\bibfnamefont {R.}~\bibnamefont {Inoue}},
		\bibinfo {author} {\bibfnamefont {Y.}~\bibnamefont {Okuyama}}, \bibinfo
		{author} {\bibfnamefont {A.}~\bibnamefont {Nakamoto}}, \ and\ \bibinfo
		{author} {\bibfnamefont {M.}~\bibnamefont {Kozuma}},\ }\href {\doibase
		10.1103/PhysRevA.91.063414} {\bibfield  {journal} {\bibinfo  {journal}
			{Physical Review A}\ }\textbf {\bibinfo {volume} {91}},\ \bibinfo {pages}
		{063414} (\bibinfo {year} {2015})}\BibitemShut {NoStop}%
	\bibitem [{\citenamefont {Yamamoto}\ \emph {et~al.}(2016)\citenamefont
		{Yamamoto}, \citenamefont {Kobayashi}, \citenamefont {Kuno}, \citenamefont
		{Kato},\ and\ \citenamefont {Takahashi}}]{Yamamoto2016}%
	\BibitemOpen
	\bibfield  {author} {\bibinfo {author} {\bibfnamefont {R.}~\bibnamefont
			{Yamamoto}}, \bibinfo {author} {\bibfnamefont {J.}~\bibnamefont {Kobayashi}},
		\bibinfo {author} {\bibfnamefont {T.}~\bibnamefont {Kuno}}, \bibinfo {author}
		{\bibfnamefont {K.}~\bibnamefont {Kato}}, \ and\ \bibinfo {author}
		{\bibfnamefont {Y.}~\bibnamefont {Takahashi}},\ }\href {\doibase
		10.1088/1367-2630/18/2/023016} {\bibfield  {journal} {\bibinfo  {journal}
			{New Journal of Physics}\ }\textbf {\bibinfo {volume} {18}},\ \bibinfo
		{pages} {023016} (\bibinfo {year} {2016})}\BibitemShut {NoStop}%
	\bibitem [{\citenamefont {Brown}\ \emph {et~al.}(2017)\citenamefont {Brown},
		\citenamefont {Mitra}, \citenamefont {Guardado-Sanchez}, \citenamefont
		{Schau{\ss}}, \citenamefont {Kondov}, \citenamefont {Khatami}, \citenamefont
		{Paiva}, \citenamefont {Trivedi}, \citenamefont {Huse},\ and\ \citenamefont
		{Bakr}}]{Brown2017}%
	\BibitemOpen
	\bibfield  {author} {\bibinfo {author} {\bibfnamefont {P.~T.}\ \bibnamefont
			{Brown}}, \bibinfo {author} {\bibfnamefont {D.}~\bibnamefont {Mitra}},
		\bibinfo {author} {\bibfnamefont {E.}~\bibnamefont {Guardado-Sanchez}},
		\bibinfo {author} {\bibfnamefont {P.}~\bibnamefont {Schau{\ss}}}, \bibinfo
		{author} {\bibfnamefont {S.~S.}\ \bibnamefont {Kondov}}, \bibinfo {author}
		{\bibfnamefont {E.}~\bibnamefont {Khatami}}, \bibinfo {author} {\bibfnamefont
			{T.}~\bibnamefont {Paiva}}, \bibinfo {author} {\bibfnamefont
			{N.}~\bibnamefont {Trivedi}}, \bibinfo {author} {\bibfnamefont {D.~A.}\
			\bibnamefont {Huse}}, \ and\ \bibinfo {author} {\bibfnamefont {W.~S.}\
			\bibnamefont {Bakr}},\ }\href {\doibase 10.1126/science.aam7838} {\bibfield
		{journal} {\bibinfo  {journal} {Science}\ }\textbf {\bibinfo {volume}
			{357}},\ \bibinfo {pages} {1385} (\bibinfo {year} {2017})}\BibitemShut
	{NoStop}%
	\bibitem [{SI()}]{SI}%
	\BibitemOpen
	\href@noop {} {\bibinfo  {journal} {See Supplemental Material below}\ }\BibitemShut {NoStop}%
	\bibitem [{\citenamefont {Parsons}\ \emph {et~al.}(2015)\citenamefont
		{Parsons}, \citenamefont {Huber}, \citenamefont {Mazurenko}, \citenamefont
		{Chiu}, \citenamefont {Setiawan}, \citenamefont {Wooley-Brown}, \citenamefont
		{Blatt},\ and\ \citenamefont {Greiner}}]{Parsons2015}%
	\BibitemOpen
	\bibfield  {journal} {  }\bibfield  {author} {\bibinfo {author} {\bibfnamefont
			{M.~F.}\ \bibnamefont {Parsons}}, \bibinfo {author} {\bibfnamefont
			{F.}~\bibnamefont {Huber}}, \bibinfo {author} {\bibfnamefont
			{A.}~\bibnamefont {Mazurenko}}, \bibinfo {author} {\bibfnamefont {C.~S.}\
			\bibnamefont {Chiu}}, \bibinfo {author} {\bibfnamefont {W.}~\bibnamefont
			{Setiawan}}, \bibinfo {author} {\bibfnamefont {K.}~\bibnamefont
			{Wooley-Brown}}, \bibinfo {author} {\bibfnamefont {S.}~\bibnamefont {Blatt}},
		\ and\ \bibinfo {author} {\bibfnamefont {M.}~\bibnamefont {Greiner}},\ }\href
	{\doibase 10.1103/PhysRevLett.114.213002} {\bibfield  {journal} {\bibinfo
			{journal} {Physical Review Letters}\ }\textbf {\bibinfo {volume} {114}},\
		\bibinfo {pages} {213002} (\bibinfo {year} {2015})}\BibitemShut {NoStop}%
	\bibitem [{\citenamefont {Greif}\ \emph {et~al.}(2016)\citenamefont {Greif},
		\citenamefont {Parsons}, \citenamefont {Mazurenko}, \citenamefont {Chiu},
		\citenamefont {Blatt}, \citenamefont {Huber}, \citenamefont {Ji},\ and\
		\citenamefont {Greiner}}]{Greif2016}%
	\BibitemOpen
	\bibfield  {author} {\bibinfo {author} {\bibfnamefont {D.}~\bibnamefont
			{Greif}}, \bibinfo {author} {\bibfnamefont {M.~F.}\ \bibnamefont {Parsons}},
		\bibinfo {author} {\bibfnamefont {A.}~\bibnamefont {Mazurenko}}, \bibinfo
		{author} {\bibfnamefont {C.~S.}\ \bibnamefont {Chiu}}, \bibinfo {author}
		{\bibfnamefont {S.}~\bibnamefont {Blatt}}, \bibinfo {author} {\bibfnamefont
			{F.}~\bibnamefont {Huber}}, \bibinfo {author} {\bibfnamefont
			{G.}~\bibnamefont {Ji}}, \ and\ \bibinfo {author} {\bibfnamefont
			{M.}~\bibnamefont {Greiner}},\ }\href {\doibase 10.1126/science.aad9041}
	{\bibfield  {journal} {\bibinfo  {journal} {Science}\ }\textbf {\bibinfo
			{volume} {351}},\ \bibinfo {pages} {953} (\bibinfo {year}
		{2016})}\BibitemShut {NoStop}%
	\bibitem [{\citenamefont {Parsons}\ \emph {et~al.}(2016)\citenamefont
		{Parsons}, \citenamefont {Mazurenko}, \citenamefont {Chiu}, \citenamefont
		{Ji}, \citenamefont {Greif},\ and\ \citenamefont {Greiner}}]{Parsons2016}%
	\BibitemOpen
	\bibfield  {author} {\bibinfo {author} {\bibfnamefont {M.~F.}\ \bibnamefont
			{Parsons}}, \bibinfo {author} {\bibfnamefont {A.}~\bibnamefont {Mazurenko}},
		\bibinfo {author} {\bibfnamefont {C.~S.}\ \bibnamefont {Chiu}}, \bibinfo
		{author} {\bibfnamefont {G.}~\bibnamefont {Ji}}, \bibinfo {author}
		{\bibfnamefont {D.}~\bibnamefont {Greif}}, \ and\ \bibinfo {author}
		{\bibfnamefont {M.}~\bibnamefont {Greiner}},\ }\href {\doibase
		10.1126/science.aag1430} {\bibfield  {journal} {\bibinfo  {journal}
			{Science}\ }\textbf {\bibinfo {volume} {353}},\ \bibinfo {pages} {1253}
		(\bibinfo {year} {2016})}\BibitemShut {NoStop}%
	\bibitem [{\citenamefont {Khatami}\ and\ \citenamefont
		{Rigol}(2011)}]{Khatami2011}%
	\BibitemOpen
	\bibfield  {author} {\bibinfo {author} {\bibfnamefont {E.}~\bibnamefont
			{Khatami}}\ and\ \bibinfo {author} {\bibfnamefont {M.}~\bibnamefont
			{Rigol}},\ }\href {\doibase 10.1103/PhysRevA.84.053611} {\bibfield  {journal}
		{\bibinfo  {journal} {Physical Review A}\ }\textbf {\bibinfo {volume} {84}},\
		\bibinfo {pages} {053611} (\bibinfo {year} {2011})}\BibitemShut {NoStop}%
	\bibitem [{\citenamefont {Fubini}\ \emph {et~al.}(2007)\citenamefont {Fubini},
		\citenamefont {Falci},\ and\ \citenamefont {Osterloh}}]{Fubini2007}%
	\BibitemOpen
	\bibfield  {author} {\bibinfo {author} {\bibfnamefont {A.}~\bibnamefont
			{Fubini}}, \bibinfo {author} {\bibfnamefont {G.}~\bibnamefont {Falci}}, \
		and\ \bibinfo {author} {\bibfnamefont {A.}~\bibnamefont {Osterloh}},\ }\href
	{\doibase 10.1088/1367-2630/9/5/134} {\bibfield  {journal} {\bibinfo
			{journal} {New Journal of Physics}\ }\textbf {\bibinfo {volume} {9}},\
		\bibinfo {pages} {134} (\bibinfo {year} {2007})}\BibitemShut {NoStop}%
	\bibitem [{\citenamefont {Cincio}\ \emph {et~al.}(2009)\citenamefont {Cincio},
		\citenamefont {Dziarmaga}, \citenamefont {Meisner},\ and\ \citenamefont
		{Rams}}]{Cincio2009}%
	\BibitemOpen
	\bibfield  {author} {\bibinfo {author} {\bibfnamefont {L.}~\bibnamefont
			{Cincio}}, \bibinfo {author} {\bibfnamefont {J.}~\bibnamefont {Dziarmaga}},
		\bibinfo {author} {\bibfnamefont {J.}~\bibnamefont {Meisner}}, \ and\
		\bibinfo {author} {\bibfnamefont {M.~M.}\ \bibnamefont {Rams}},\ }\href
	{\doibase 10.1103/PhysRevB.79.094421} {\bibfield  {journal} {\bibinfo
			{journal} {Physical Review B}\ }\textbf {\bibinfo {volume} {79}},\ \bibinfo
		{pages} {094421} (\bibinfo {year} {2009})}\BibitemShut {NoStop}%
	\bibitem [{\citenamefont {Hoang}\ \emph {et~al.}(2016)\citenamefont {Hoang},
		\citenamefont {Bharath}, \citenamefont {Boguslawski}, \citenamefont {Anquez},
		\citenamefont {Robbins},\ and\ \citenamefont {Chapman}}]{Hoang2016}%
	\BibitemOpen
	\bibfield  {author} {\bibinfo {author} {\bibfnamefont {T.~M.}\ \bibnamefont
			{Hoang}}, \bibinfo {author} {\bibfnamefont {H.~M.}\ \bibnamefont {Bharath}},
		\bibinfo {author} {\bibfnamefont {M.~J.}\ \bibnamefont {Boguslawski}},
		\bibinfo {author} {\bibfnamefont {M.}~\bibnamefont {Anquez}}, \bibinfo
		{author} {\bibfnamefont {B.~A.}\ \bibnamefont {Robbins}}, \ and\ \bibinfo
		{author} {\bibfnamefont {M.~S.}\ \bibnamefont {Chapman}},\ }\href {\doibase
		10.1073/pnas.1600267113} {\bibfield  {journal} {\bibinfo  {journal} {PNAS}\
		}\textbf {\bibinfo {volume} {113}},\ \bibinfo {pages} {9475} (\bibinfo {year}
		{2016})}\BibitemShut {NoStop}%
	\bibitem [{\citenamefont {Kantian}\ \emph {et~al.}()\citenamefont {Kantian},
		\citenamefont {Langer},\ and\ \citenamefont {Daley}}]{Kantian2016}%
	\BibitemOpen
	\bibfield  {author} {\bibinfo {author} {\bibfnamefont {A.}~\bibnamefont
			{Kantian}}, \bibinfo {author} {\bibfnamefont {S.}~\bibnamefont {Langer}}, \
		and\ \bibinfo {author} {\bibfnamefont {A.~J.}\ \bibnamefont {Daley}},\
	}\href {\doibase 10.1103/PhysRevLett.120.060401} {\bibfield
	{journal} {\bibinfo  {journal} {Physical Review Letters}\ }\textbf {\bibinfo
		{volume} {120}},\ \bibinfo {pages} {060401} (\bibinfo {year}
	{2018})}\BibitemShut {NoStop}%
	\bibitem [{\citenamefont {Lee}\ \emph {et~al.}(2006)\citenamefont {Lee},
		\citenamefont {Nagaosa},\ and\ \citenamefont {Wen}}]{Lee2006}%
	\BibitemOpen
	\bibfield  {author} {\bibinfo {author} {\bibfnamefont {P.~A.}\ \bibnamefont
			{Lee}}, \bibinfo {author} {\bibfnamefont {N.}~\bibnamefont {Nagaosa}}, \ and\
		\bibinfo {author} {\bibfnamefont {X.~G.}\ \bibnamefont {Wen}},\ }\href
	{\doibase 10.1103/RevModPhys.78.17} {\bibfield  {journal} {\bibinfo
			{journal} {Reviews of Modern Physics}\ }\textbf {\bibinfo {volume} {78}},\
		\bibinfo {pages} {17} (\bibinfo {year} {2006})}\BibitemShut {NoStop}%
	\bibitem [{\citenamefont {Mazloom}\ \emph {et~al.}(2017)\citenamefont
		{Mazloom}, \citenamefont {Vermersch}, \citenamefont {Baranov},\ and\
		\citenamefont {Dalmonte}}]{Mazloom2017}%
	\BibitemOpen
	\bibfield  {author} {\bibinfo {author} {\bibfnamefont {A.}~\bibnamefont
			{Mazloom}}, \bibinfo {author} {\bibfnamefont {B.}~\bibnamefont {Vermersch}},
		\bibinfo {author} {\bibfnamefont {M.~A.}\ \bibnamefont {Baranov}}, \ and\
		\bibinfo {author} {\bibfnamefont {M.}~\bibnamefont {Dalmonte}},\ }\href
	{\doibase 10.1103/PhysRevA.96.033602} {\bibfield  {journal} {\bibinfo
			{journal} {Physical Review A}\ }\textbf {\bibinfo {volume} {96}},\ \bibinfo
		{pages} {033602} (\bibinfo {year} {2017})}\BibitemShut {NoStop}%
	\bibitem [{\citenamefont {Zhang}\ and\ \citenamefont {Duan}(2013)}]{Zhang2013}%
	\BibitemOpen
	\bibfield  {author} {\bibinfo {author} {\bibfnamefont {Z.}~\bibnamefont
			{Zhang}}\ and\ \bibinfo {author} {\bibfnamefont {L.-M.}\ \bibnamefont
			{Duan}},\ }\href {\doibase 10.1103/PhysRevLett.111.180401} {\bibfield
		{journal} {\bibinfo  {journal} {Physical Review Letters}\ }\textbf {\bibinfo
			{volume} {111}},\ \bibinfo {pages} {180401} (\bibinfo {year}
		{2013})}\BibitemShut {NoStop}%
	\bibitem [{\citenamefont {Farhi}\ \emph {et~al.}(2001)\citenamefont {Farhi},
		\citenamefont {Goldstone}, \citenamefont {Gutmann}, \citenamefont {Lapan},
		\citenamefont {Lundgren},\ and\ \citenamefont {Preda}}]{Farhi2001}%
	\BibitemOpen
	\bibfield  {author} {\bibinfo {author} {\bibfnamefont {E.}~\bibnamefont
			{Farhi}}, \bibinfo {author} {\bibfnamefont {J.}~\bibnamefont {Goldstone}},
		\bibinfo {author} {\bibfnamefont {S.}~\bibnamefont {Gutmann}}, \bibinfo
		{author} {\bibfnamefont {J.}~\bibnamefont {Lapan}}, \bibinfo {author}
		{\bibfnamefont {A.}~\bibnamefont {Lundgren}}, \ and\ \bibinfo {author}
		{\bibfnamefont {D.}~\bibnamefont {Preda}},\ }\href {\doibase
		10.1126/science.1057726} {\bibfield  {journal} {\bibinfo  {journal}
			{Science}\ }\textbf {\bibinfo {volume} {292}},\ \bibinfo {pages} {472}
		(\bibinfo {year} {2001})}\BibitemShut {NoStop}%
\end{thebibliography}

\begin{thebibliography}{3}%
	\makeatletter
	\providecommand \@ifxundefined [1]{%
		\@ifx{#1\undefined}
	}%
	\providecommand \@ifnum [1]{%
		\ifnum #1\expandafter \@firstoftwo
		\else \expandafter \@secondoftwo
		\fi
	}%
	\providecommand \@ifx [1]{%
		\ifx #1\expandafter \@firstoftwo
		\else \expandafter \@secondoftwo
		\fi
	}%
	\providecommand \natexlab [1]{#1}%
	\providecommand \enquote  [1]{``#1''}%
	\providecommand \bibnamefont  [1]{#1}%
	\providecommand \bibfnamefont [1]{#1}%
	\providecommand \citenamefont [1]{#1}%
	\providecommand \href@noop [0]{\@secondoftwo}%
	\providecommand \href [0]{\begingroup \@sanitize@url \@href}%
	\providecommand \@href[1]{\@@startlink{#1}\@@href}%
	\providecommand \@@href[1]{\endgroup#1\@@endlink}%
	\providecommand \@sanitize@url [0]{\catcode `\\12\catcode `\$12\catcode
		`\&12\catcode `\#12\catcode `\^12\catcode `\_12\catcode `\%12\relax}%
	\providecommand \@@startlink[1]{}%
	\providecommand \@@endlink[0]{}%
	\providecommand \url  [0]{\begingroup\@sanitize@url \@url }%
	\providecommand \@url [1]{\endgroup\@href {#1}{\urlprefix }}%
	\providecommand \urlprefix  [0]{URL }%
	\providecommand \Eprint [0]{\href }%
	\providecommand \doibase [0]{http://dx.doi.org/}%
	\providecommand \selectlanguage [0]{\@gobble}%
	\providecommand \bibinfo  [0]{\@secondoftwo}%
	\providecommand \bibfield  [0]{\@secondoftwo}%
	\providecommand \translation [1]{[#1]}%
	\providecommand \BibitemOpen [0]{}%
	\providecommand \bibitemStop [0]{}%
	\providecommand \bibitemNoStop [0]{.\EOS\space}%
	\providecommand \EOS [0]{\spacefactor3000\relax}%
	\providecommand \BibitemShut  [1]{\csname bibitem#1\endcsname}%
	\let\auto@bib@innerbib\@empty
	\bibitem [{\citenamefont {Parsons}\ \emph {et~al.}(2016)\citenamefont
	{Parsons}, \citenamefont {Mazurenko}, \citenamefont {Chiu}, \citenamefont
	{Ji}, \citenamefont {Greif},\ and\ \citenamefont {Greiner}}]{Parsons2016a}%
\BibitemOpen
\bibfield  {author} {\bibinfo {author} {\bibfnamefont {M.~F.}\ \bibnamefont
	{Parsons}}, \bibinfo {author} {\bibfnamefont {A.}~\bibnamefont {Mazurenko}},
\bibinfo {author} {\bibfnamefont {C.~S.}\ \bibnamefont {Chiu}}, \bibinfo
{author} {\bibfnamefont {G.}~\bibnamefont {Ji}}, \bibinfo {author}
{\bibfnamefont {D.}~\bibnamefont {Greif}}, \ and\ \bibinfo {author}
{\bibfnamefont {M.}~\bibnamefont {Greiner}},\ }\href {\doibase
10.1126/science.aag1430} {\bibfield  {journal} {\bibinfo  {journal}
	{Science}\ }\textbf {\bibinfo {volume} {323}},\ \bibinfo {pages} {1253}
(\bibinfo {year} {2016})}\BibitemShut {NoStop}%

\bibitem [{\citenamefont {Mazurenko}\ \emph {et~al.}(2017)\citenamefont
	{Mazurenko}, \citenamefont {Chiu}, \citenamefont {Ji}, \citenamefont
	{Parsons}, \citenamefont {Kan{\'{a}}sz-Nagy}, \citenamefont {Schmidt},
	\citenamefont {Grusdt}, \citenamefont {Demler}, \citenamefont {Greif},\ and\
	\citenamefont {Greiner}}]{Mazurenko2017a}%
\BibitemOpen
\bibfield  {author} {\bibinfo {author} {\bibfnamefont {A.}~\bibnamefont
		{Mazurenko}}, \bibinfo {author} {\bibfnamefont {C.~S.}\ \bibnamefont {Chiu}},
	\bibinfo {author} {\bibfnamefont {G.}~\bibnamefont {Ji}}, \bibinfo {author}
	{\bibfnamefont {M.~F.}\ \bibnamefont {Parsons}}, \bibinfo {author}
	{\bibfnamefont {M.}~\bibnamefont {Kan{\'{a}}sz-Nagy}}, \bibinfo {author}
	{\bibfnamefont {R.}~\bibnamefont {Schmidt}}, \bibinfo {author} {\bibfnamefont
		{F.}~\bibnamefont {Grusdt}}, \bibinfo {author} {\bibfnamefont
		{E.}~\bibnamefont {Demler}}, \bibinfo {author} {\bibfnamefont
		{D.}~\bibnamefont {Greif}}, \ and\ \bibinfo {author} {\bibfnamefont
		{M.}~\bibnamefont {Greiner}},\ }\href {\doibase 10.1038/nature22362}
	{\bibfield  {journal} {\bibinfo  {journal} {Nature}\ }\textbf {\bibinfo
	{volume} {545}},\ \bibinfo {pages} {462} (\bibinfo {year}
{2017})}\BibitemShut {NoStop}%

	\bibitem [{\citenamefont {Z{\"{u}}rn}\ \emph {et~al.}(2013)\citenamefont
		{Z{\"{u}}rn}, \citenamefont {Lompe}, \citenamefont {Wenz}, \citenamefont
		{Jochim}, \citenamefont {Julienne},\ and\ \citenamefont {Hutson}}]{Zurn2013}%
	\BibitemOpen
	\bibfield  {author} {\bibinfo {author} {\bibfnamefont {G.}~\bibnamefont
			{Z{\"{u}}rn}}, \bibinfo {author} {\bibfnamefont {T.}~\bibnamefont {Lompe}},
		\bibinfo {author} {\bibfnamefont {A.~N.}\ \bibnamefont {Wenz}}, \bibinfo
		{author} {\bibfnamefont {S.}~\bibnamefont {Jochim}}, \bibinfo {author}
		{\bibfnamefont {P.~S.}\ \bibnamefont {Julienne}}, \ and\ \bibinfo {author}
		{\bibfnamefont {J.~M.}\ \bibnamefont {Hutson}},\ }\href {\doibase
		10.1103/PhysRevLett.110.135301} {\bibfield  {journal} {\bibinfo  {journal}
			{Physical Review Letters}\ }\textbf {\bibinfo {volume} {110}},\ \bibinfo
		{pages} {135301} (\bibinfo {year} {2013})}\BibitemShut {NoStop}%
	\bibitem [{\citenamefont {Khatami}\ and\ \citenamefont
		{Rigol}(2011)}]{Khatami2011a}%
	\BibitemOpen
	\bibfield  {author} {\bibinfo {author} {\bibfnamefont {E.}~\bibnamefont
			{Khatami}}\ and\ \bibinfo {author} {\bibfnamefont {M.}~\bibnamefont
			{Rigol}},\ }\href {\doibase 10.1103/PhysRevA.84.053611} {\bibfield  {journal}
		{\bibinfo  {journal} {Physical Review A}\ }\textbf {\bibinfo {volume} {84}},\
		\bibinfo {pages} {053611} (\bibinfo {year} {2011})}\BibitemShut {NoStop}%
\end{thebibliography}
\end{document}